\input epsf

\documentclass[12pt]{article}

\usepackage{paper2e}
\usepackage{mydefs2e}

\usepackage{epsfig}

\renewcommand{\Box}{\,\raisebox{-.45pt}{\drawsquare{7}{0.6}}\,}
\newcommand{\grad}{{\vec\nabla}}
\newcommand{\MP}{M_{\rm Pl}}

\renewcommand{\d}{\partial}
\newcommand{\nab}[1]{\nabla\!_{#1}}

\newcommand{\pc}{\phi_{\rm c}}

\begin{document}
\begin{titlepage}

\preprint{HUTP-05/A0030\\UTAP-530\\RESCEU-10/05}

\title{Dynamics of Gravity in a Higgs Phase}

\author{Nima Arkani-Hamed$^{\rm a}$, Hsin-Chia Cheng$^{\rm a,b}$,
Markus A. Luty$^{\rm a,c,d}$,\\ Shinji Mukohyama$^{\rm a,e}$,
Toby Wiseman$^{\rm a}$}

\address{$^{\rm a}$Jefferson Laboratory of Physics, Harvard University\\
Cambridge, Massachusetts 02138}

\address{$^{\rm b}$Department of Physics, University of California\\
Davis, California 95616}

\address{$^{\rm c}$Physics Department, Boston University\\
Boston, Massachusetts 02215}

\address{$^{\rm d}$Physics Department, University of Maryland\\
College Park, Maryland 20742$\,$\footnote{Permanent address}}

\address{$^{\rm e}$Department of Physics and 
Research Center for the Early Universe\\
The University of Tokyo, Tokyo 113-0033, Japan}

\begin{abstract}
We investigate the universal low-energy dynamics of the simplest Higgs
phase for gravity, `ghost condensation.'
We show that the nonlinear dynamics of the `ghostone' field dominate
for all interesting gravitational sources.
Away from caustic singularities, the dynamics is equivalent to the
irrotational flow of a perfect fluid with equation of state
$p \propto \rho^2$, where the fluid particles can have negative mass.
We argue that this theory is free from catastrophic instabilities due to
growing modes, even though the null energy condition is violated.
Numerical simulations show that solutions generally have
singularities in which
negative energy regions shrink to zero size.
We exhibit partial UV completions of the theory in which
these singularities are smoothly resolved, so this does not
signal any inconsistency in the effective theory.
We also consider the bounds on the symmetry breaking scale
$M$ in this theory.
We argue that the nonlinear dynamics cuts off the Jeans
instability of the linear theory, and allows $M \lsim 100\GeV$.
\end{abstract}

\end{titlepage}

\section{Introduction}
Is general relativity the correct description of gravity at long
distances and times?
Certainly, there are good reasons for thinking that this is the case.
Experimentally, gravity has been probed at distance scales ranging
from $10^{-1}$~mm (in short-range force experiments) to at least
$10^{14}$~cm
(the size of the solar system).
Theoretically, general relativity is the unique Lorentz-invariant
theory of massless spin 2, and its conceptual elegance is beyond
question.
However, the situation is far less clear on cosmological distance and time
scales.
Structure formation, galaxy rotation curves and gravitational lensing,
and the accelerating expansion of the universe cannot be explained by
general relativity coupled to known matter.
These anomalous effects are conventionally attributed to `dark
matter' and `dark energy.'
However, given the fact that the observed effects
are purely gravitational, it makes sense to ask whether they may
have a common origin in a modification of gravity in the infrared.
These considerations have led to a revival of interest in consistent
infrared modifications of 
gravity~\cite{Gregory:2000jc,Dvali:2000hr,Arkani-Hamed:2002sp,Carroll:2003wy,Arkani-Hamed:2003uy,Rubakov:2004eb,Bekenstein:2004ca}.

In the present paper, we further investigate the model of \Ref{Arkani-Hamed:2003uy},
`ghost condensation.'
This can be viewed as the universal low-energy dynamics associated
with the simplest Higgs phase for gravity, arising when Lorentz symmetry
is broken spontaneously.
Breaking of Lorentz symmetry is of course ubiquitous.
For example, time-dependent fields in cosmology define a preferred frame.
However, these solutions are not the ground state of the theory:
they carry energy density and dilute away as the universe expands.
Any `modification of gravity' induced by such solutions becomes relevant
only at scales of order the horizon.
We are instead interested in the situation where Lorentz symmetry is
broken in flat spacetime, allowing nontrivial modification of gravity
inside the horizon.
This means that the symmetry breaking sector has peculiar properties;
in particular, the stress-energy tensor must vanish in the ground
state:
\beq
\avg{T_{\mu\nu}} = 0.
\eeq

Spontaneous breaking of Lorentz symmetry gives rise to a gapless scalar
excitation analogous to the Nambu-Goldstone bosons that arise from
spontaneous breaking of internal symmetries.
`Ghost condensation' gives rise to a single such mode, and is in this
sense the minimal model of spontaneous breaking of Lorentz symmetry.
We refer to the scalar mode as a `ghostone boson.'
In analogy with the Higgs phase for gauge theory, the ghostone mode
mixes with the graviton, modifying gravity in a nontrivial manner.

\Ref{Arkani-Hamed:2003uy} studied this theory and analyzed the modification of gravity
in the weak-field limit.
The dynamics is governed by a consistent effective theory defined by
the scale $M$ where the symmetry is broken.
It was shown that the ghostone mode gives a 
possible new origin for dark energy and dark matter.
\Ref{Arkani-Hamed:2003uz} showed that the ghostone mode may also be the inflaton,
leading to interesting testable
consequences. 

In the present paper, we further investigate the dynamics of this theory.
We show that nonlinearities dominate the dynamics of the 
ghostone sector for all gravitational sources of interest.
In particular, the time scale for the onset of nonlinear dynamics for
a fixed gravitational source is precisely the infall (or orbit) time
associated with the source.
Away from singularities, the nonlinear solutions are equivalent to the
gradient flow of a fluid with equation of state
\beq
p = \frac{\rho^2}{2 M^4}.
\eeq
The gradient fluid flow picture breaks down at caustic singularities where
geodesics of fluid particles cross.
At these points higher-derivative terms become important.
Numerical simulations show that these remove the caustic singularity
in most cases, but in some cases singularities still form.
These correspond to regions of negative energy that shrink in size until
their dynamics becomes sensitive to the underlying microscopic theory.
We argue that these can be stabilized in a partial UV completion of
the model into `gauged ghost condensation.'

We also address stability issues in the nonlinear theory.  Although
the energy is not positive-definite in the nonlinear theory, there are
no growing modes corresponding to infinite positive energy radiated to
infinity.  We argue that the worst instability is that of negative
energy regions shrinking in size, as found in the simulations.

Based on these results, we address the question of the bounds on the
ghostone sector.  We find that the most sensitive bounds come from the
fact that the Jeans instability found in \Ref{Arkani-Hamed:2003uy}
produces regions of positive and negative energy and that those regions
induce scintillation of light rays coming from far distances by
gravitational lensing. Demanding that the total angular deviation by the
scintillation, or the random walk, not exceed the observed angular
resolution gives the bound $M \lsim 100\GeV$.  Other possible observable
effects such as modification of gravity near astrophysical sources and
energy loss of moving sources are argued to be negligibly small.  Black
holes in the ghost condensate are studied in \Ref{Mukohyama:2005rw},
where it is found that they also do not give strong constraints on the
model (see also \Ref{Frolov}).


This paper is organized as follows.
In section 2, we review the Higgs phase for gravity,
emphasizing the universal nature of the low-energy dynamics.
We also review the origin of the relevant length and time scales
in the linearized theory.
In section 3, we discuss the nonlinear dynamics of the ghost condensate.
We discuss the time scales and present the fluid picture.
In section 4, we show the numerical simulations of the nonlinear evolutions 
and discuss the resolution of caustic singularities.
In section 5, we discuss the bounds of this theory.
This includes mass and energy accretion in slow-moving objects,
gravitational lensing and energy loss from moving objects.
We do not claim to have a complete understanding of the dynamics,
so this section is intended to be preliminary and provocative.
In section 6, we briefly discuss the possibility that ghost condensate
may constitute the dark matter. We show that the initial growth of the
density perturbations in the linear regime is identical to that of the
standard cold dark matter. Whether it can form the correct structure
depends on the details of the nonlinear evolution which is left for future
investigations.  
Our conclusions are presented in section 7.

\section{Review of the Linear Theory}

\subsection{Effective Theory}
What is a Higgs phase for gravity?
It is easiest to answer this question in linearized general relativity,
where we expand the metric about flat space
\beq
g_{\mu\nu} = \eta_{\mu\nu} + h_{\mu\nu}
\eeq
and keep only terms quadratic in $h_{\mu\nu}$.
In this theory, the fields $h_{\mu\nu}$ are closely analogous to
gauge fields with gauge transformation law
\beq
\de h_{\mu\nu} = -(\d_\mu \xi_\nu + \d_\nu \xi_\mu),
\eeq
where $\xi^\mu$ are the generators of infinitesmal diffeomorphisms
\beq
x^\mu \mapsto x^\mu + \xi^\mu.
\eeq

We want to consider the case where the time diffeomorphisms
generated by $\xi^0$ are spontaneously broken.
This means that time diffeomorphisms are realized nonlinearly
in the effective theory containing only the ghostone field.
The minimal model contains a single real ghostone field $\pi$
that shifts under time diffeomorphisms:
\beq
\de \pi = -\xi_0.
\eeq
Note that $\pi$ naturally has units of time.
We now write the most general effective Lagrangian invariant under these symmetries.
This contains the Einstein Lagrangian, and additional terms constructed from the
invariants
\beq
\Si &= \dot\pi - \sfrac 12 h_{00},
\\
K_{ij} &= \frac{1}{2}(\dot{h}_{ij} - \d_i h_{0j} - \d_j h_{0i} + 2 \d_i \d_j \pi).
\eeq
The leading terms are
\beq[Leffpilinear]
\scr{L}_{\rm eff} = \scr{L}_{\rm E}
+ M^4 \left\{
\frac 12 \left( \dot{\pi} - \sfrac 12 h_{00} \right)^2
- \frac{\al_1}{2M^2} K_{ij}^2
- \frac{\al_2}{2M^2} K^2 + \scr{O}(\pi^3) \right\}.
\eeq
(Note that we have built in the fact that flat space is a solution by
not writing any linear terms in the Lagrangian.)

In the limit where we turn off gravity, we see that the ghostone
mode has dispersion relation
\beq
\om^2 = \frac{\al \vec{k}^4}{M^2},
\eeq
where $\al = \al_1 + \al_2$.
If the dimensionless couplings in the effective
Lagrangian are order 1,
the mass scale $M$ is the scale of new physics in this theory.

\Eq{Leffpilinear} is the starting point for analyzing the linear dynamics
of the theory, including the coupling to gravity.
However, there are several points that may not be completely clear
in this formulation.
First, it may not be completely clear in what sense the breaking of
Lorentz symmetry is spontaneous, since the time direction appears to be
explicitly singled out.
Second, it is not clear whether the absence of a linear term in $\Si$
(allowed by all symmetries) requires an additional fine-tuning, like
the cosmological constant which is a linear term proportional to $h^\mu_\mu$.
Third, we would like to have an efficient way to extend this to the full
nonlinear theory.
All of these issues can be elegantly understood in the covariant formulation,
which we turn to next.

\subsection{Covariant Formulation}
Consider an effective theory with a real scalar $\phi$ that is invariant
under a global shift symmetry
\beq
\de \phi = \la,
\eeq
where $\la$ is a spacetime independent constant.
For example, $\phi$ could be a Nambu-Goldstone boson from the breaking of a global
symmetry, or a 0-form gauge field in string theory.

A conventional effective Lagrangian for this theory is
\beq
\scr{L} = +\sfrac 12 \d^\mu \phi \d_\mu \phi
\eeq
with equation of motion $\Box \phi = 0$.
This has solutions with $\phi = n_\mu x^\mu$ for any constant
4-vector $n_\mu$.
If $n_\mu$ is timelike, we can choose the time direction so that
the solution is
\beq
\phi = c t.
\eeq
This is a solution for any constant $c$.
At first sight it may appear that these are obviously not candidate
ground states of the theory, but the situation is actually more subtle.
Suppose we expand in fluctuations about this solution
\beq[phifluct]
\phi = c t + \pi.
\eeq
Note that under time diffeomorphisms, $\pi$ transforms as
\beq
\de \pi = -\xi_0/c,
\eeq
just like the ghostone mode considered above.
Expanding the Lagrangian to quadratic order in $\pi$, one
finds that the fluctuations for $\pi$ have good
time and space kinetic terms.
This means that the solution is stable under local fluctuations
for any value of $c\,$!
The reason for this is that the theory has a conserved current
associated with the shift symmetry
\beq
J_\mu = \d_\mu \phi.
\eeq
Solutions with $c \ne 0$ have a constant nonzero charge density.
Local excitations cannot change the total charge, so configurations
with lower energy cannot be reached.
However, when we turn on gravity, solutions with $c \ne 0$ will
cause the universe to expand, and the charge will dilute away.
Lorentz invariance is therefore not broken spontaneously in this
theory.

Consider instead an effective Lagrangian of the form
\beq
\scr{L}_{\rm eff} = M^4 P(X),
\qquad
X = \d^\mu \phi \d_\mu \phi.
\eeq
Note that $\phi$ has dimensions of length (or time), so that $X$ is dimensionless.
This omits only terms with more than one derivative acting on
$\phi$, such as $(\Box\phi)^2$.
We will include their effects below.
This also has solutions of the form $\phi = c t$ for any $c$.
Expanding in fluctuations \Eq{phifluct}, we obtain
\beq
\scr{L}_{\rm eff} = M^4 \left\{ 
\left[ 2 c^2 P''(c^2) + P'(c^2) \right] \dot\pi^2 -
P'(c^2) (\grad \pi)^2 + \scr{O}(\pi^3) \right\}
\eeq
We see that small perturbations are stable provided
\beq
2 c^2 P''(c^2) + P'(c^2) > 0,
\qquad
P'(c^2) > 0.
\eeq
(Note that a conventional kinetic term $P(X) = + \frac 12 X$
satisfies these conditions.)

\begin{figure}
 \centering\leavevmode\epsfysize=5cm \epsfbox{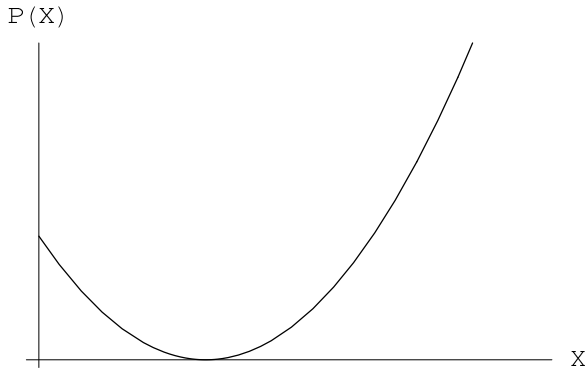}
 \caption{\label{fig:kin1} The kinetic fuction $P(X)$ for ghost condensation.}
\end{figure}

To see whether any of these solutions may be regarded as candidate
ground states, we compute the stress-energy tensor:
\beq
T_{\mu\nu} &= M^4 \left[
-P(X) g_{\mu\nu} + 2 P'(X) \d_\mu \phi \d_\nu \phi \right].
\eeq
In the solution $\phi = c t$, the first term is proportional to $g_{\mu\nu}$
and can be cancelled by tuning the cosmological constant.
The second term gives rise to
\beq
T_{00} = 2 c^2 P'(c^2),
\qquad
T_{ij} = 0.
\eeq
This gives rise to an expanding universe unless
\beq
P'(c^2) = 0.
\eeq
Note that the conserved shift current is given by
\beq
J_\mu = 2 P'(X) \d_\mu \phi,
\eeq
so the charge density vanishes in this configuration.  Therefore, if
$P(X)$ has the form shown in Fig.~\ref{fig:kin1}, then the
configuration with $P'(X) = 0$ is a candidate ground state.  This is
the `ghost condensate.'  From now on, we rescale the field $\phi$ so
that the ground state is at $X = 1$, and write
\beq
\Si = \sfrac 12 (X - 1).
\eeq

Stability in the linearized theory requires higher-derivative terms to give
a nonzero spatial kinetic term, for example
\beq
\De\scr{L}_{\rm eff} = -\frac{\al M^2}{2} (\Box \phi)^2 
= -\frac{\al M^2}{2} (\grad^2 \pi)^2 + \cdots.
\eeq
This leads to the $\om^2 \sim \vec{k}^4$ dispersion relation for the
ghostone mode already discussed above.

\subsection{Higgs Phase for Gravity}

The ghostone mode mixes with gravity, giving rise to a nontrivial
modification of gravity in the infrared.
A detailed analysis is given in \Ref{Arkani-Hamed:2003uy}, but much of the physics
can be understood from very general considerations, which we review
here.

\begin{figure}
 \centering\leavevmode\epsfysize=8cm \epsfbox{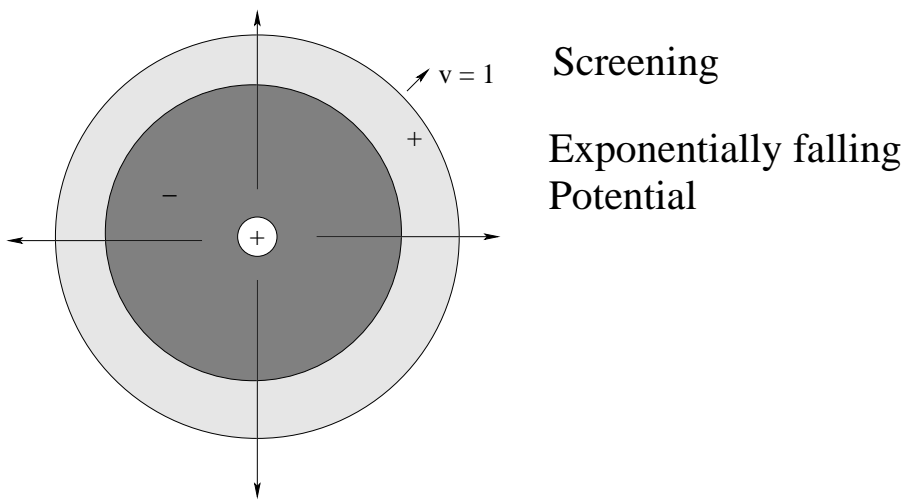}
 \caption{\label{fig:gaugehiggs}A schematic illustration of the screening
  effect for gauge theories in the Higgs phase.}
\end{figure}

Recall first the ordinary Higgs mechanism for gauge theory.
In this case, a nonzero charge gives rise to
a tadpole for the Goldstone mode.
Excitations of the Goldstone mode themselves carry charge, and because
opposite charges attract, the Goldstone charge tends to screen the
original charge.
If the charge configuration is time-dependent, the Goldstone field
responds at the speed of light.
This is illustrated schematically in Fig.~\ref{fig:gaugehiggs}.

\begin{figure}
 \centering\leavevmode\epsfysize=8cm \epsfbox{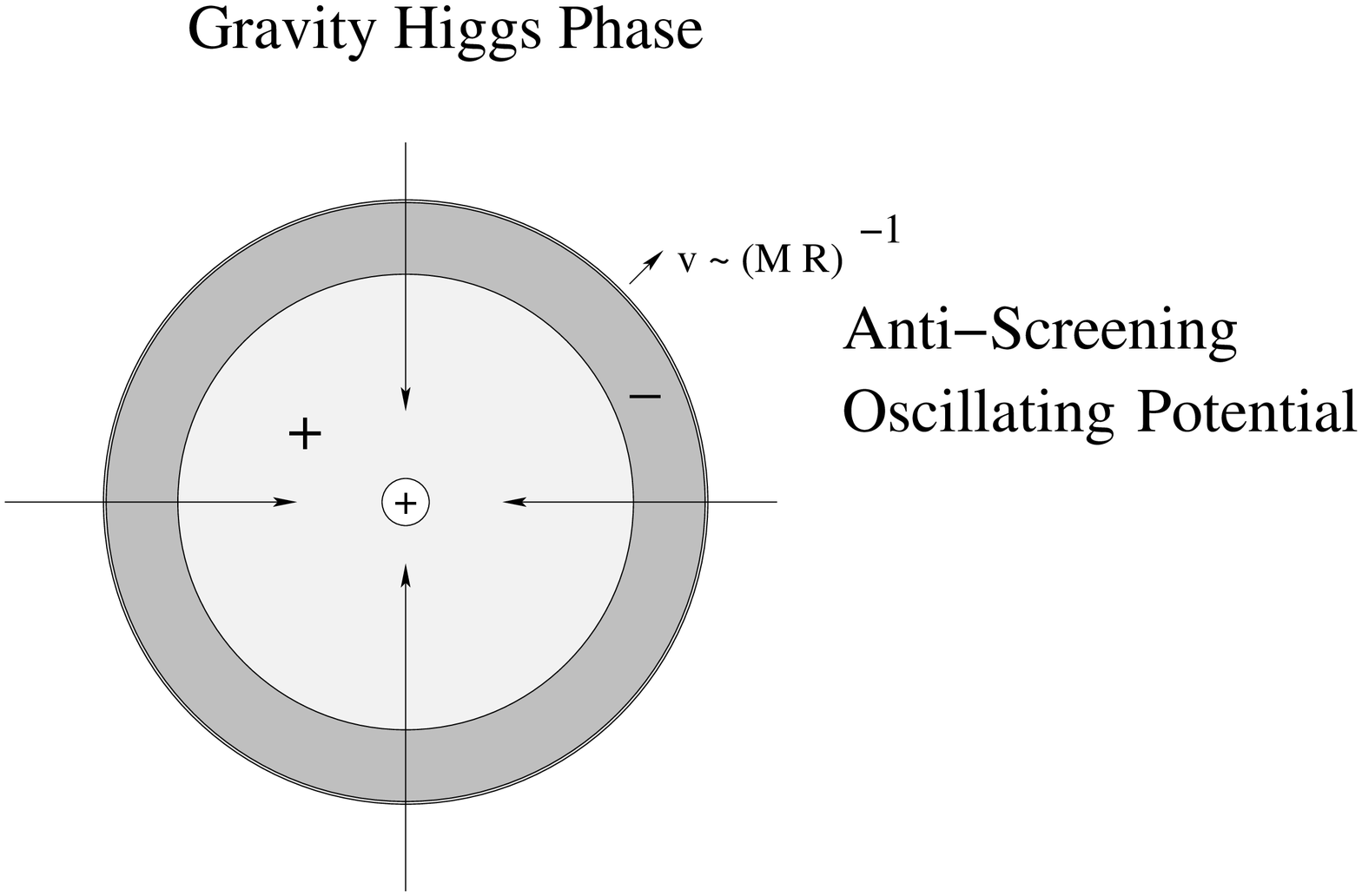}
 \caption{\label{fig:gravhiggs}A schematic illustration of the anti-screening
  effect for gravity in the Higgs phase}
\end{figure}

Similarly, in the Higgs mechanism for gravity, a nonzero mass (the charge
that sources gravity) gives rise to a tadpole for the ghostone mode.
Excitations of the ghostone mode themselves carry energy, but because
positive masses attract in gravity, the ghostone field \emph{anti}-screens
the source gravitational field.
This gives rise to an instability of the vacuum analogous to the Jeans
instability of pressureless matter coupled to gravity.
The ghostone Jeans instability can be seen in the dispersion
relation for the scalar mode that follows from the mixing with gravity:
\beq
\om^2 = \frac{\al \vec{k}^4}{M^2} - \frac{\al M^2}{2\MP^2} \vec{k}^2.
\eeq
This is illustrated schematically in Fig.~\ref{fig:gravhiggs}.
The length and time scale associated with the Jeans
instability are
\beq[LJ-TJ]
L_{\rm J} \sim \frac{\MP}{M^2},
\qquad
T_{\rm J} \sim \frac{\MP^2}{M^3} \gg L_{\rm J}.
\eeq
For $M \lsim 10\MeV$, $T_{\rm J}$ is longer than the lifetime of
the universe and there is clearly no constraint from the Jeans
instability.

\subsection{Negative Energy}
Even for time shorter than the Jeans time scale $T_{\rm J}$,
stability is an issue because the ghostone energy can be negative.
Recall that the ground state $X = 1$ is the boundary of the stability region.
Expanding to higher orders in $\pi$, we find interaction terms such as
\beq[Lintcubic]
\scr{L}_{\rm eff} = M^4 \left[ \sfrac 12 \dot\pi^2
- \sfrac 12 \dot\pi (\grad \pi)^2 + \cdots 
-\frac{\al}{2M^2} (\grad^2 \pi)^2 \right].
\eeq
In regions where $\dot\pi < 0$ the cubic term gives rise to
negative gradient energy, and therefore an instability.
Classically, the $(\grad^2\pi)^2$ term restores stability at short
wavelengths, and there is a growing mode only for long wavelengths:
\beq
\la \gsim \frac{M^{-1}}{\avg{\dot\pi}^{1/2}}.
\eeq
Quantum-mechanically, there are fluctuations at all length scales.
Nonetheless, \Ref{Arkani-Hamed:2003uy} showed that there is no quantum
instability in the effective theory using a scaling argument.  If we
scale energy by $E \to s E$, the quadratic kinetic terms are left
invariant if we scale
\beq
\bal
t &\to s^{-1} t,
\\
\vec{x} &\to s^{-1/2} \vec{x},
\\
\pi &\to s^{1/4} \pi.
\eal
\eeq
With this scaling the cubic operator $\dot\pi (\grad \pi)^2$ 
in \Eq{Lintcubic} scales
as $s^{1/4}$ and is therefore (barely!) irrelevant.
All other operators are even more irrelevant, showing that there
is a regime of low energies where the expansion is under control.

\subsection{\label{subsec:preview}Preview of Nonlinear Effects}

The arguments above show that the effects of the nonlinear terms are
under control at low energies and small field amplitudes.
However, in the presence of large classical gravitational sources
the nonlinear terms can become important.
In fact, the time scale for the ghostone field near a classical source
is just the gravitational infall time of the source.
This can be understood from the form of the stress-energy tensor.
In the approximation where the Lagrangian is $\scr{L} = P(X)$, 
the stress-energy tensor has the form
\beq
T_{\mu\nu} \propto 
-P(X) g_{\mu\nu} + 2 P'(X) u_\mu  u_\nu  .
\eeq
where
\beq
u_\mu = \d_\mu \phi.
\eeq
This has the form of a stress-energy tensor for a perfect fluid with
4-velocity $u_\mu$.%
\footnote{The defining property of a perfect fluid is that at each point
there is a frame in which the stress-energy tensor has the form
$T_{\mu\nu} = \diag(-\rho, p, p, p)$.}
The fact that $u_\mu$ is a gradient means that the flow is irrotational:
$\d_{[\mu} u_{\nu]} = 0$ implies $\grad \times \vec{u} = 0$.
The equations of motion for the ghostone field follow from the conservation
of the stress-energy tensor $\nabla^\mu T_{\mu\nu} = 0$,
which are intepreted as the conservation of energy and momentum in the fluid.
In the presence of a classical gravitational source, a fluid clearly
responds on a time scale given by the infall time, and therefore
so does the ghostone mode. As it will be shown in the next section,
this is exactly the time scale where the nonlinear term becomes important.

For small fluctuations about the minimum $X=1$, $P(X)$ can be approximated
by 
\beq
P(X) \approx \frac{1}{8}(X-1)^2 = \frac{1}{2} \Sigma^2,
\eeq
where an overall contribution to the cosmological constant has been removed.
We can read off
the equation of state of the fluid from the energy-momentum tensor.
It is
\beq
p = \frac{\rho^2}{2 M^4},
\eeq
where
\beq
\rho = M^4 \Si.
\eeq
For $\rho \ll M^4$, small ghostone fluctuations therefore behave
like an almost pressureless fluid, which makes the ghostone excitations
a candidate for the dark matter in our universe.
We will have more to say on this later.

As we noted above, Hubble expansion drives $\Si \to 0$,
and therefore $\rho \to 0$.
Even for $\rho = 0$ the fluid has nontrivial dynamics in the
presence of gravitational sources: the fluid particles follow
geodesics.
As a result, the local
preferred frame (where $u_\mu = \d_\mu \phi$ is
`at rest') is a freely-falling frame.
This frame is also the preferred frame from the spontaneous
breaking of Lorentz invariance, so we conclude that the
`aether' is dragged by the local
gravitational field.

Since geodesics generically cross (especially in the presence of
gravitational sources), the field configuration develops caustic
singularities where the field gradients diverge.
Such phenomena were also observed in some other scalar field 
theories~\cite{Felder:2002sv}.
Near these singularities, the higher-derivative
$\al(\grad^2 \pi)^2$ term can no longer be neglected.
This term contributes positive gradient energy, and we find in
numerical simulations that
it generically smoothly resolves the caustic singularity, giving rise
to a `bounce' with outgoing $\pi$ waves and positive and negative
regions of $\Si$ near the would-be caustic.

The fact that $\Si$ (and hence $\rho$) can be negative brings up
again the question of the stability of the theory.
As discussed above, regions with $\Si < 0$
modes with sufficiently long wavelengths are unstable.
In numerical simulations, we find that negative energy regions
tend to shrink while the amplitude of $\Si$ grows inside the region.
This can be understood from the fluid picture, since this is valid
in the limit where we neglect the $(\grad^2\pi)^2$ term,
which is a good approximation away from caustic singularities.
In this picture the $\Si < 0$ region
consists of fluid particles with negative mass.
It is therefore clear that energy (mass)
cannot flow across the boundary of the $\Si < 0$ region, since
the boundary consists of particles with vanishing mass.
The boundary can move however, and in the $\Si < 0$ region the
positive pressure favors large gradients
and causes the $\Si < 0$ region to shrink.
Numerical simulations show that some $\Si < 0$ regions
continue to shrink until they exit the regime of validity of the
effective theory.
These singularties need to be resolved in a more fundamental theory.
Similar conclusion was also obtained in \Ref{Krotov:2004if} which studied 
the two-dimensional case.
However, we show that the total energy inside the $\Si < 0$ regions
formed in astrophysical situations is very small, and does not
lead to any observable consequences provided the singularities
are regulated in a smooth way.
We will discuss a partial UV completion to do this, and present
numerical evidence that it works.

The nonlinear dynamics affects the bounds on $M$ derived in
the linear theory. 
The Jeans instability in the linear theory gives
a bound $M \lsim 10$ MeV if we require that there is no exponential growth 
of the oscillatory potential within the age of the universe.
However, the nonlinear dynamics is expected to cut off the instability,
and may weaken this bound.
We argue below that the strongest bound comes from gravitational lensing
due to regions of positive and negative energy produced by the Jeans
instability. Demanding that the random walk of light rays due to the
lensing does not smear out the observed CMB anisotropies gives the bound
$M \lsim 100\GeV$.

We also consider other possible bounds on the ghost condensate from
the gravitational sector.  The modifications of the gravitational
potential are small due to velocity effects
\cite{Dubovsky:2004qe,Peloso:2004ut}.  We also consider energy loss in
the nonlinear theory, as well as energy stored in would-be caustic
singularities.  We find that these effects are negligible, and we
believe that the theory is safe for $M \lsim 100\GeV$.


\section{Nonlinear Dynamics}
We now turn to the nonlinear dynamics of the theory.
The nonlinear dynamics is very rich and complex, and we emphasize
that we do not claim
a complete understanding in this work.
It is therefore important to keep in mind that
there is a simple limit of this theory, independent of the details of
the nonlinear dynamics \cite{Arkani-Hamed:2003uy}.
The Ghostone sector naturally couples to matter only through gravity.
(Gravitationally induced direct couplings to standard model fields are easily
seen to be negligibly small.)
The maximum value of the gravitational energy in the Ghostone energy is
of order $M^4$, which does not affect even cosmology if
$M \lsim (\MP H_0)^{1/2} \sim 10^{-3}\eV$.
Such low values of $M$ are still very interesting for cosmology, since the ghost
may be a source of dark energy and dark matter \cite{Arkani-Hamed:2003uy} and may drive
inflation \cite{Arkani-Hamed:2003uz}.

Another general point to keep in mind in the following is that the
modifications of gravity vanish in regions where $\Si = 0$ ($X = 1$)
and we neglect the $(\grad^2\pi)^2$ term.
This is because in this limit, the Lagrangian in `unitary gauge'
$\phi = t$ ($\phi = 0$) is
\beq
\scr{L} = \sqrt{-g} P(g^{00}).
\eeq
This has the form of a gauge-fixing term for the gauge $X = g^{00} = 1$,
so if $X = 1$ initially, the gravitational dynamics is unchanged.
This means that any modification of gravity will be suppressed by
the small amplitude $\Si$ or by the $\al(\grad^2 \pi)^2$ term which
is small at long wavelengths.
We will see that this suppresses many possible effects compared to
\naive expectations.

In the remainder of the paper,
we consider $M \gg 10^{-3}\eV$ to see what modifications
of gravity may be observable today, and determine
the experimental limit on the scale $M$.

\subsection{Effective Lagrangian}
We are interested in the gravitational dynamics of large, slow-moving sources,
where relativistic and retardation effects are negligible.
We further restrict ourselves to the domain of weak gravity, so we are
not considering black holes.
(Black holes in the presence of the ghost condensate are discussed in \Ref{Mukohyama:2005rw}.)
It is convenient to incorporate these approximations from the beginning
in an effective Lagrangian.
The fact that gravity is weak means that we use the linearized
approximation for gravity.
The fact that the system is nonrelativistic means that $\d_t \ll \grad$.
For the motions of objects, we can neglect gravitational radiation,
so we can parameterize the metric by the scalar potential $\Phi$ as
\beq
ds^2 = (1 + 2 \Phi) dt^2 - (1 - 2 \Phi) d\vec{x}^{\,2},
\eeq
where
\beq
\grad^2 \Phi = \frac{T_{00}}{2\MP^2}.
\eeq
We also assume that ghostone amplitudes are small in units of $M$.
Since only derivatives of $\pi$ are meaningful, this means that
\beq
| \grad \pi | \ll 1.
\eeq
With these approximations, the deviation of $X$ from its minimum
is given by
\beq[Siform]
\Si = \sfrac 12 (X - 1) = \dot\pi - \sfrac{1}{2} (\grad \pi)^2 - \Phi.
\eeq
Note that there is nothing in the assumptions above that determines the
relative size of the terms on the \rhs.
The effective Lagrangian is then
\beq[Leffnonlinear]
\scr{L}_{\rm eff} = M^4 \left\{
\sfrac 12 \Si^2 - \frac{\al}{2M^2} (\grad^2 \pi)^2 \right\}.
\eeq
All the terms omitted are smaller than the ones we have kept by the assumptions
above.
Note that all the nonlinear effects in this approximation are due to the
$(\grad \pi)^2$ term in $\Si$.

\subsection{Nonlinear Time Scale}
\label{subsec:nonlineartimescale}
We now estimate the time scale for the nonlinear gradient terms in
\Eq{Siform} to dominate the dynamics.
We assume that the gravitational potential $\Phi$ is determined by external sources,
and varies on a length scale $L$ in the region of interest.
(For example, at a distance $r$ outside a gravitational source,
$L \sim r$.)
If initially $\pi = 0$, the gravitational potential gives a tadpole
that forces $\pi$ to be nonzero.
We want to know the
time scale $T_{\rm NL}$ when the nonlinear term in $\Si$ becomes as important
as the linear $\dot\pi$ term.
This is determined by
\beq[Sidet]
\frac{\pi}{T_{\rm NL}} \sim \frac{\pi^2}{L^2} \sim \Phi,
\eeq
where $\pi$ is a typical ghostone amplitude.
This gives
\beq
T_{\rm NL} \sim \frac{L}{\sqrt\Phi}.
\eeq
We recognize this as the gravitational response time associated with the
potential $\Phi$.
For example, for a point source $\Phi \sim R_{\rm S} / r$ where $R_{\rm S}$
is the Schwarzschild radius, and we have
$T_{\rm NL}^2 \sim r^{3/2} / R_{\rm S}$,
which is the Kepler relation.
This is a very direct way of seeing that the nonlinear effects become important
on the gravitational time scale.

Solving \Eq{Sidet} for $\pi$, we obtain
\beq
| \grad \pi| \sim \frac{\pi}{L} \sim \sqrt{\Phi} \ll 1.
\eeq
That is, ghostone amplitudes are small for weak gravity,
in agreement with our assumptions.

On the other hand, in the linear approximation 
the $\al(\grad^2 \pi)^2$ becomes comparable to $\dot{\pi}^2$
at $T_{Lin} \sim M L^2$. As a result, the
nonlinear evolutions completely dominates for $T_{NL} \lsim T_{Lin} \sim M L^2$, or
\beq
\Phi L^2 \gsim \frac{1}{M^2}.
\eeq
For example, outside a gravitational source of mass
$M_{\rm src}$, this condition is
\beq
r \gsim \frac{1}{M_{\rm src}} \left( \frac{\MP}{M} \right)^2,
\eeq
and the earth's surface gravity is in the nonlinear regime for
$M \gsim 10^{-8}\eV$.

As we have emphasized, the modification of gravity is suppressed
by $\Si$ or by the $\al(\grad^2 \pi)^2$ term.
However, we will argue below that with the nonlinear evolution, 
these also become important at the
time scale $T_{\rm NL}$ found above because this is the timescale for
the formation of caustic singularities.
This is to be compared with the time scale for the modification of gravity
in the linear regime, which is the time scale for the Jeans instability.
If we take $M \lsim 10\MeV$ (as in \Ref{Arkani-Hamed:2003uy}),
the Jeans time scale is longer than the age
of the universe, and the gravitational time scale of \emph{any} 
object is a much shorter time scale. 
We see that the nonlinear effects completely dominate the dynamics
in all regimes of interest.%
%

\subsection{Fluid Picture}
We now show that the nonlinear equations of motion have a simple interpretation
in terms of a perfect fluid if we neglect the $\al(\grad^2 \pi)^2$ term
in the action.
As already mentioned in subsection \ref{subsec:preview}, we can see this connection
directly in terms of the stress-energy tensor for the ghostone mode.
Near $X = 1$ we can approximate $P(X)$ by
\beq
P(X) = \sfrac 18 (X - 1)^2 = \sfrac 12 \Si^2.
\eeq
Neglecting the $\al(\grad^2 \pi)^2$ term, the stress-energy tensor has the form
\beq
T_{\mu\nu} = M^4 \left\{ -\sfrac 12 \Si^2 g_{\mu\nu} + \Si\, u_\mu u_\nu \right\},
\eeq
where
\beq
u_\mu = \d_\mu \phi.
\eeq
Note that $u_\mu$ is nonzero and timelike everywhere.
This has the form of the stress-energy tensor for a perfect fluid with
4-velocity $u_\mu$.
Because $u_\mu$ is the gradient of a scalar, the flow of the fluid is
irrotational.
The conservation of the stess-energy tensor $\nabla^\mu T_{\mu\nu} = 0$
gives the equation of motion for the ghostone field, and also gives the
Euler equation for the fluid.
For $\Si \ll 1$ we can read off the density and pressure
\beq[equofstate]
\rho = M^4 \Si,
\qquad
p = \sfrac 12 M^4 \Si^2 = \frac{\rho^2}{2M^4}.
\eeq
This establishes the equivalence of the ghostone theory without the
$\al(\grad^2 \pi)^2$ term to the irrotational flow of a perfect
fluid~\footnote{
The Jeans length $\tilde{L}_J=2\pi c_s/\sqrt{4\pi G_N\rho}$ 
corresponding to this equation of state is $\tilde{L}_J\sim M_{Pl}/M^2$,
where $c_s=\sqrt{dp/d\rho}=\sqrt{\rho/M^4}$ is the sound 
velocity. Intriguingly, this agrees with the Jeans length
(\ref{eq:LJ-TJ}) in the linear theory up to a constant of order
unity. This equation of state ignores the $k^4$ term, while the linear
analysis does not take into account the nonlinear term $\Sigma^2$ in
$p$. Hence, it is not a priori clear whether these two Jeans length
should be the same or not. Nonetheless, they agree.}.

It is also insightful to understand the equivalence directly in terms
of the equations of motion.
For the ghostone field, the equations of motion
have the form of a conservation law
\beq[eom1]
\dot\Si = \grad \cdot [\Si \grad \pi],
\eeq
where $\Si$ is the charge density.
For a fluid made of particles carrying the conserved charge, the current is
$\vec{J} = \Si \vec{v}$, so we identify
\beq
\vec v = -\grad \pi.
\eeq
Again we see that the fluid flow is irrotational.
In this fluid picture, the equations of motion are satisfied simply
due to the fact that the fluid particles carry their charges with them.
It remains only to satisfy the relation between $\Si$ and $\pi$:
\beq[vinterp]
\Si = \dot\pi - \sfrac 12 (\grad \pi)^2 - \Phi.
\eeq
Taking the gradient of both sides and using \Eq{vinterp} gives
\beq[fluidparteom]
\frac{D \vec{v}}{D t} = -\grad(\Phi + \Si),
\eeq
where
\beq
\frac{D}{Dt} = \frac{\d}{\d t} + \vec{v} \cdot \grad
\eeq
is the time derivative along the particle worldline (also called the
convective or Lagrangian derivative).
\Eq{fluidparteom} is just Newton's law for a particle moving in a potential
$\Phi + \Si$.
Using the identifications \Eq{equofstate} we can write
\beq
-\grad\Si = -\frac{1}{\rho} \grad p,
\eeq
which shows that \Eq{fluidparteom} is precisely Euler's equation for
the fluid.%
\footnote{Note that the equivalence between the
ghostone and fluid pictures is a kind of duality, since it exchanges a
constraint equation with an equation of motion.}

The equivalence between the ghostone dynamics and the dynamics of a perfect
fluid was derived here in the nonrelativistic approximation and for linearized
gravity.
In fact, as shown in the appendix,
it holds in full nonlinear general relativity.

We can now understand better the justification for neglecting the
$\al(\grad^2 \pi)^2$ term, which was used to derive the fluid picture.
As discussed in the introduction, outside a gravitational source, the fluid
(and therefore the ghostone field) responds on a time scale of order the
gravitational infall time of the source.
For reasonable values of $M$, this is much shorter than the time for
the $\al(\grad^2 \pi)^2$ term to become important.
However, the particle trajectories in a perfect fluid tend to cross,
leading to caustic singularities.
(For $\Si \ne 0$, the pressure may stop these singularities, but they
certainly occur for $\Si = 0$.)
At the caustic singularity, derivatives of the velocity field blow up,
and higher-derivative terms such as the $\al(\grad^2 \pi)^2$ term cannot
be neglected.
Therefore, the fluid picture is valid away from small regions where
caustics form.

We can also understand better the extent to which the locally preferred
frame (the `aether') is dragged by the gravitational field.
We expect the universe to consist of domains where $\Si$ is negligbly
small and $\phi$ varies smoothy, separated by caustic regions.
Away from the caustics, the locally preferred frame is a freely-falling
frame, and in this sense the aether is dragged by the local gravitational
field.

\subsection{Negative Energy}
\label{subsec:negativeenergy}
We now re-examine the question of negative energy in the theory in the
presence of the nonlinearities discussed above.
Because the time translation invariance and $\phi$ shift invariance are
spontaneously broken down to a diagonal time shift symmetry, there are
different possible definitions of energy in this theory.
Gravity couples to the Noether charge of the original time translation
symmetry, which we refer to as the `gravitational energy.'
This is not the same as the Noether
charge associated with the time translation symmetry that is unbroken in the
vacuum, which we call the `inertial energy.'

The inertial energy $\varepsilon$
is the energy associated with the unbroken time translation
symmetry of the Lagrangian \Eq{Leffnonlinear}.
(It is also the Hamiltonian density of the system.)
Conservation of inertial energy states that
\beq
\dot\varepsilon = -\grad \cdot \vec{p},
\eeq
where
\beq[rhonl]
\varepsilon = M^4 \left\{ \sfrac 12 \Si^2
+ \sfrac 12 \Si (\grad \pi)^2
+ \frac{\al}{2 M^2} (\grad^2 \pi)^2 \right\}
\eeq
is the inertial energy density and
\beq
\vec{p} = M^4 \left\{
-\Si \dot\pi \grad\pi + \frac{\al}{M^2} \left[
(\grad^2 \pi) \grad \pi  - \dot\pi \grad (\grad^2 \pi) \right]
\right\}
\eeq
is the momentum density.
In the linearized approximation,
\beq
\varepsilon = M^4 \left\{ \sfrac 12 \dot\pi^2 + \frac{\al}{2M^2}
(\grad^2 \pi)^2 \right\} \ge 0.
\eeq
However, in the nonlinear theory the inertial energy is not positive definite
due to the second term in \Eq{rhonl}.
The energy density can be negative only in regions where $\Si < 0$.
In fact it is easy to see that the energy is unbounded from below.
For example, for $\pi = c |\vec x|$ we have $\varepsilon = -\sfrac 18 M^4 c^4$.

We might expect this theory to be unstable because any region can lower
its energy by emitting positive energy radiation to infinity.
(Because inertial energy is positive in the linear theory, radiation must have
positive energy.)
The reason this does not happen is that there is another conserved
quantity in the theory, namely the Noether charge associated with the
$\phi$ shift symmetry.%
\footnote{The gravitational energy is a linear combination of the inertial energy
and the shift charge, so the conservation of the stress-energy tensor
does not give an independent conservation law.}
Because $\pi$ shifts under the shift symmetry, the conservation of shift
charge is identical to the equation of motion for $\pi$:
\beq
\dot \Si = -\grad \cdot \vec J,
\eeq
where $\Si$ is the shift charge density, and the shift current is
\beq
\vec J = -\Si \grad \pi + \frac{\al}{M^2} \grad(\grad^2 \pi).
\eeq

If we neglect the $\al(\grad^2 \pi)^2$ term, this conservation
law is taken into account very directly in the fluid picture, to which we
now turn.
In this approximation, negative energy regions correspond precisely to
regions where $\Si < 0$, so we consider such a region surrounded by
$\Si > 0$.%
\footnote{We do not consider configurations that have $\Si < 0$ at infinity.
These can be eliminated by physical initial conditions.
For example, in the expanding universe Hubble friction drives can drive
$\Si \to 0$ from above.}
In the fluid picture, the conserved charge
is carried by the individual particles, so the $\Si < 0$ region consists
of particles with negative charge, while the
boundary of the $\Si < 0$ region consists of particles of vanishing charge.
Therefore, there can be no flux of charge across the $\Si = 0$ boundary,
and the total charge inside the region does not change.

On the other hand, the $\Si = 0$ boundary can move.
Since $\grad\Si$ points outward at the boundary, the equation of
motion for fluid particles \Eq{fluidparteom} implies that the
particles on the boundary experience an inward force due to
the pressure.
Therefore, the $\Si < 0$ region tends to shrink.

These arguments show that the total shift charge integrated over a
$\Si < 0$ region
\beq
Q = \int_{\Si < 0} d^3 x\, \Si
\eeq
does not change with time.
This can also be seen from the fact that the shift current
$\vec{J}$ vanishes on the $\Si = 0$ boundary.
The shift charge is not the same as the total inertial energy
\beq
\scr{E} = \int_{\Si < 0} d^3 x\, \varepsilon.
\eeq
However, if we neglect the $\al (\grad^2 \pi)^2$ term, the flux of
inertial energy across the $\Si = 0$ boundary also vanishes, and
therefore $\scr{E}$ also does not change with time.

As with the fluid picture, these results hold beyond the approximations
made here.
This is discussed in the appendix.

\subsection{\label{subsec:causticsoln}Caustic Solutions}
The fluid picture can be used to understand the structure of the caustic
singularties that occur when we neglect the $\al(\grad^2 \pi)^2$ term.
We restrict attention to the case $\Si = 0$, where there is no pressure
and the fluid particles follow geodesics.
In this case, it is clear that there are caustics without the
$\al(\grad^2 \pi)^2$ term.
It is possible that the pressure resolves the caustics in important
situations such as inside galaxy halos made of ghostone dark matter,
but we will not consider that here.

To introduce the subject of caustics, we consider the gravitational
potential due to a uniform sphere of matter with density $\rho_0$.
Inside the sphere, the gravitational potential is
\beq
\Phi = \frac{\rho_0}{6\MP^2} r^2,
\eeq
\ie\ a harmonic potential.
If we consider an initial fluid configuration where all particles
are at rest, then all fluid particles reach the center of the sphere
at the same time!
This kind of `perfect caustic' requires very special initial
conditions and symmetry.
Nonetheless, we will see that an analog of the perfect caustic
appears to be responsible for the singularities that remain
in the full theory even after the $\al(\grad^2 \pi)^2$ term is included.

A more realistic situation will have less symmetry in the gravitational
potential and the initial conditions.
Nonetheless, in the presence of gravitational sources the fluid particle
trajectories tend to cross, resulting in caustic singularities.
The generic situation is that families of trajectories cross on a
dimension-2 surface.
We want to zoom in on the behavior near the caustic
surface, so we can treat the problem as 1-dimensional,
and approximate the caustic surface by a plane.
We also expect the potential to be irrelevant on the small scales
where the caustic structure is important.
This reduces the problem to a very simple 1-dimensional one.

In the absence of a potential, the particle trajectories are just
straight lines:
\beq[lines]
x = x_0 + v_0(x_0) t.
\eeq
Here we are labeling each trajectory by its initial position $x_0$
at $t = 0$.
The initial velocity function $v_0(x_0)$ determines the geodesics.
The problem now is to find the velocity of the particle at a given
point $(x, t)$, which is a simple kinematic problem.

To get a caustic near $x = 0$, we are interested in the case where
the particle trajectories that start near $x_0 = 0$
are focussed toward each other:
\beq
v_0(0) = 0,
\qquad
v_0'(0) < 0.
\eeq
For small $x_0$ we can therefore approximate $v_0(x_0)$ by a linear function
of $x_0$:
\beq[v0expandperfect]
v_0(x_0) = -\frac{x_0}{T},
\eeq
where $T$ is a constant that tells us how the initial velocity
varies away from $x_0 = 0$.
Solving \Eq{lines} for $x_0$ we obtain
\beq
x_0 = -T \frac{x}{t - T}.
\eeq
Since the velocity is constant along any trajectory, we have
\beq
v(x) = v_0(x_0) = -\frac{x}{T - t}.
\eeq
We can relate this to the field $\pi$ using $\d_x \pi = -v$,
to obtain
\beq
\pi(x, t) = -\frac{x^2}{2(t - T)}.
\eeq
Note that this solution is scale-free,
since $T$ just gives the time
to the caustic.
Shifting $t \to t - T$, we obtain
\beq
\pi(x, t) = -\frac{x^2}{2t}.
\eeq

\begin{figure}
 \centering\leavevmode\epsfbox{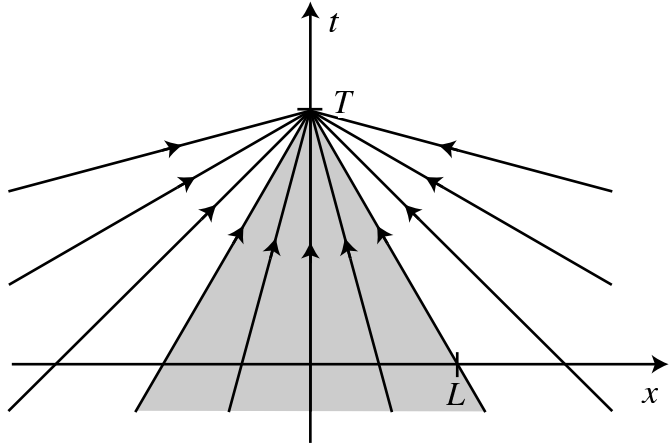}
 \caption{\label{fig:caustic} The 1-dimensional `perfect caustic.'
The shaded regions shows the region where we can expand the solution
perturbatively about the perfect caustic.}
\end{figure}

To understand this solution, note that the general
condition for a caustic is
\beq
\frac{\d x}{\d x_0} = 0
\eeq
for fixed $t$.
This gives the time to the caustic for a given value of $x_0$ as
\beq
t_{\rm c}(x_0) = -\frac{1}{v_0'(x_0)}.
\eeq
For $v_0$ given by \Eq{v0expandperfect} above,
we see that the caustic forms at $t = T$ for
all initial points $x_0$.
We call this the `perfect caustic,' and it is
illustrated in Fig.~\ref{fig:caustic}.

In a realistic case, the function $v_0(x_0)$ will be more complicated.
We want to expand about the point where the caustic first forms,
\ie the minimum of $t_{\rm c}(x_0)$,
which we take to be $x_0 = 0$.
Expanded about this point, $v_0$ takes the form
\beq[v0higherorder]
v_0(x_0) = -\frac{x_0}{T} 
+ \frac{1}{6 L^2 T} x_0^3
+ \scr{O}(L^{-3}),
\eeq
where we have performed a boost so that $v_0(0) = 0$.
Because of the focussing of the geodesics, this expansion is
valid only for
\beq
|x| \ll L \left| \frac{t - T}{T} \right|.
\eeq
This region of validity is shown in Fig.~\ref{fig:caustic}.

We can now follow the same steps as for the perfect caustic,
keeping higher order terms in \Eq{v0higherorder}.
We find
\beq
x_0 &= -T \frac{x}{t - T} 
- \frac{T^3}{6 L^2} \frac{x^3 t}{(t - T)^4}
+ \scr{O}(L^{-3}).
\eeq
and therefore
\beq[vximperfect]
v(x) &= \frac{x}{t - T}
+ \frac{T^3}{6 L^2} \frac{x^3}{(t - T)^4}
+ \scr{O}(L^{-3}).
\eeq
As above, this is related to the ghostone field solution by
$\d_x \pi = -v$.

This result can be used to evaluate the effect of the $\al (\grad^2 \pi)^2$
term in the action near the caustic singularity.
In the equation of motion for $\pi$, it contributes a term proportional
to $\al \d_x^4 \pi$.
This can be thought of as a restoring force, since it tends to smooth
out the large gradients near the caustic region.
Evaluating this in the caustic solution \Eq{vximperfect} gives
\beq[grad2picaustic]
\d_x^4 \pi = -\d_x^3 v
= -\frac{T^3}{L^2} \frac{1}{(t - T)^4}
+ \scr{O}(L^{-3}).
\eeq
Note that a nonzero contribution to $\d_x^4 \pi$ arises only at
$\scr{O}(L^{-2})$.
In particular, there is no restoring force at all in the perfect
caustic solution.
We believe that this explains the singular behavior we find 
in numerical simulations below.
Since the $\d_x^4 \pi$ vanishes in the perfect caustic, it is an
exact solution to the full nonlinear equations, including the
$\al (\grad^2 \pi)^2$ term.
This solution therefore becomes singular and exits the regime of
validity of the effective theory.
Even though the perfect caustic is a very special solution, it 
does describe the generic caustic close to the singularity.
Also, the fact that the perfect caustic
is a scale-free solution may mean that it is an attractor in the full
nonlinear dynamics.

We can use \Eq{grad2picaustic} to estimate the time and distance scale
where the $\al(\grad^2 \pi)^2$ term becomes important.
This will happen when 
\beq
\al\frac{\d_x^4 \pi}{M^2} \gsim \ddot\pi.
\eeq
For a solution that varies on a time scale $T$ and distance scale
$L$, the scaling in subsection~\ref{subsec:nonlineartimescale}
shows that $\ddot\pi \sim L^2 / T^3$.
We therefore find that the $\al (\grad^2 \pi)^2$ term becomes important
for
\beq
\De t \lsim \left( \frac{\al T^3}{M L^2} \right)^{1/2},
\eeq
where $\De t = T - t$ is the time to the caustic.
The distance scale where the $\al (\grad^2 \pi)^2$ term becomes important
is therefore
\beq
\De x \sim \frac{L}{T} \De t \lsim \left( \frac{\al T}{M} \right)^{1/2}.
\eeq
Note that $\De x, \De t \gg M^{-1}$ as long as $L, T \gg M^{-1}$
and $L / T \ll 1$ (\ie the system is nonrelativistic),
so this is within the regime of validity of the effective theory.


\section{Numerical Simulations}
In this section we describe various numerical simulations of the
ghost field which allow us to understand the rather exotic features
of its non-linear evolution.

The test cases we first present assume symmetry to reduce the dynamics
to a one dimensional problem, and will see that the various symmetries
have different behaviours. We will discuss how caustics do indeed form
in the theory. As mentioned earlier, this is clear for $\alpha=0$, but
one would naively expect non-zero $\alpha$ to ameliorate this
problem. However, as we will show, this is not the case, and for
certain symmetries the `perfect' caustic remains an attractor.

As one might expect, the singular behaviour becomes less strong as
one moves from planar through axisymmetry to spherical symmetry.
Indeed without any gravitational potential we will see the planar
reduction exhibits singularities, while the spherical theory without
potential does not. However, once a gravitationally attractive
potential is added, all three symmetries become singular under
evolution of regular initial data. 

Clearly assuming symmetry can lead to unphysical behaviours, whilst
the physical situation we are ultimately interested, namely
structure formation, is expected to have very little symmetry. Hence
this section will conclude with a study of a 3-d numerical evolution
where no continuous symmetry is present and a moving gravitational
potential seeds the ghost field growth. As expected from the 1-d
examples, we again find singular caustics do develop, and
interestingly appear to take a planar form.

\subsection{One dimensional evolutions: Planar, axial and spherical symmetry}

We will initially consider the case of evolution of the ghost field
in the absence of gravitational sources, seeded instead from a local
perturbation in the ghost field itself. Then reducing to planar
symmetry $\pi = \pi(t,r)$ and we may write the equation for the
ghost decoupled from gravity in a manifestly flux conservative form,
\begin{eqnarray}
\dot{H} & = & \partial_r \left[  \Sigma + \frac{1}{2} H^2 \right] \nonumber \\
\dot{\Sigma} & = & \partial_r \left[  \Sigma H - L^2 H'' \right] ,
\label{eq:eom}
\end{eqnarray}
where $H = \partial_r \pi$ is the gradient of the ghost field, 
$L=\sqrt{\alpha}/M$, the
second line is simply the definition of $\Sigma$ and $\dot{x}, x'$ are
the time and space derivatives of $x$. We then use a Crank-Nicholson
method to evolve our initial data, which we take to be a Gaussian
profile in $\pi$
\begin{equation}
\pi(t=0) = \pi_0 e^{ - r^2 } \label{eq:init_data}
\end{equation}
with $\Sigma = 0$ initially. As discussed previously, with $L=0$ $\Sigma$
would remain zero, but the higher derivative term sources $\Sigma$.

In order to make contact with the epoch of structure formation we
wish to have moderate initial amplitudes so $\mid \pi_0 \mid < 1$
but is still of order unity.  From (\ref{eq:init_data}) we have
chosen units to have initial data with unit spatial variation. We
then wish to have $L$, the ghost length scale, to be $L \ll 1$.
Naturally numerical methods limit our ability to separate the
initial data length scale from the ghost length whilst maintaining
accuracy. However we may separate the scale sufficiently to see the
asymptotic properties of taking $L$ very small.

\begin{figure}[ht]
\centerline{\psfig{file=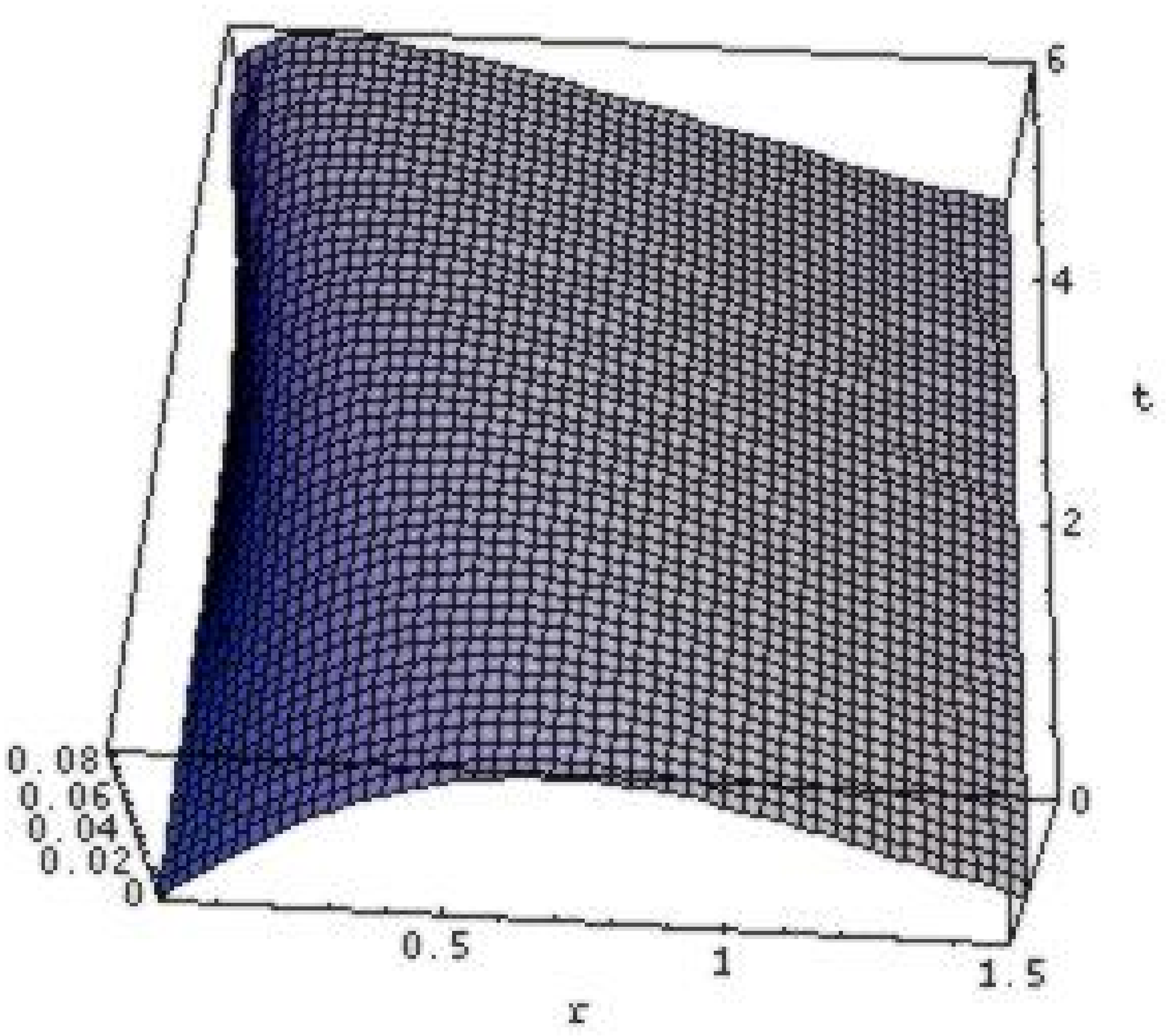,width=3.5in}
\psfig{file=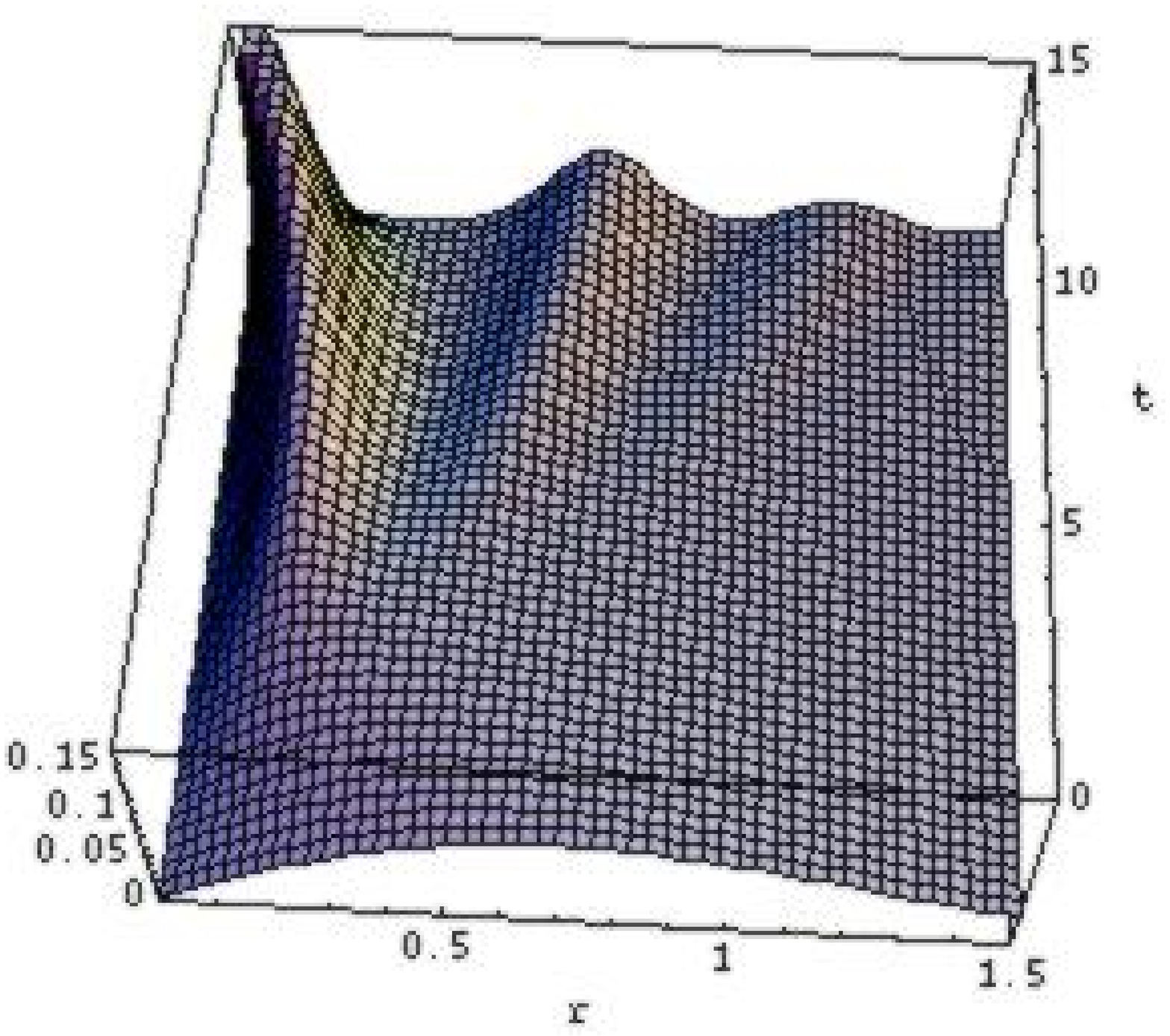,width=3.5in}}
\caption{ \label{fig:planar1} Figure on the left shows $H =
\partial_r \pi$ for planar evolution of initial Gaussian in $\pi$
with $\pi_0 = 0.1$ and $L = 0.$. Note a caustic forms as expected.
The figure on the right shows the same evolution but with non-zero
$L = 0.005$.  Both evolutions are shown up to the time where the
field becomes singular. We see the higher derivative term smooths
the initial singularity present in the $L = 0.$ case but again
results in a later singularity. }
\end{figure}

\begin{figure}[ht]
\centerline{\psfig{file=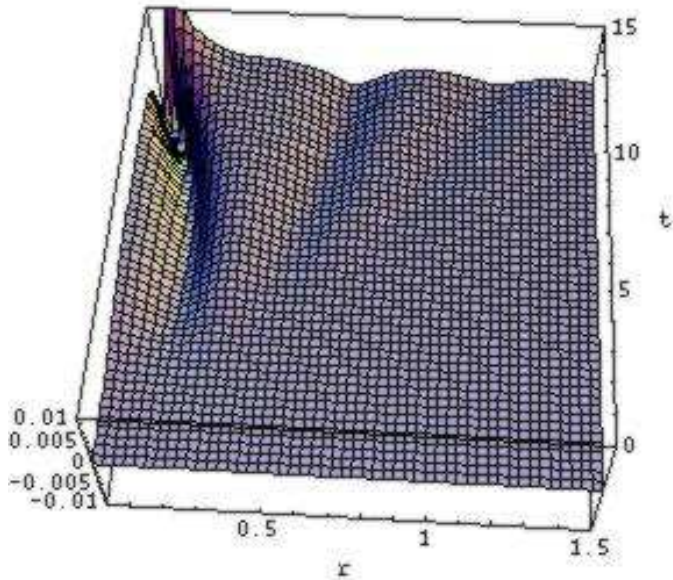,width=3.5in}}
\caption{ \label{fig:planar2} Figure showing $\Sigma$ for the same
evolution with $L = 0.005$ as in figure $\ref{fig:planar1}$.  }
\end{figure}

The first figures we show, \ref{fig:planar1} and \ref{fig:planar2},
illustrate the generic behaviour for the planar system for small
$L$, showing both $H$ and $\Sigma$ for the evolution. We show both the
evolution of data for $L = 0.005$ and also for $L = 0.$ for
comparison, and the initial amplitude was taken to be $\pi_0 = 0.1$.

As expected $L = 0.$ evolution leads to a caustic, whose time for
formation scales as $t_{\mathrm{caustic}} \simeq 1/\pi_0$. Adding
the higher derivative term changes the evolution dramatically around
this time scale, completely smoothing out the ghost field. However,
we see from the figure that the radiated waves scattered by the
action of this higher derivative term in fact are attracted back to
the origin where they grow and become singular. Obviously since the
symmetry is planar, this growth is not due to a measure factor, but
rather is due a `perfect caustic' forming, which as mentioned before
cannot be rescued by the higher derivative term we use here. Indeed
this perfect caustic formation can be seen locally in detail from
$\Sigma$ near the singularity. 

\begin{figure}[ht]
\centerline{\psfig{file=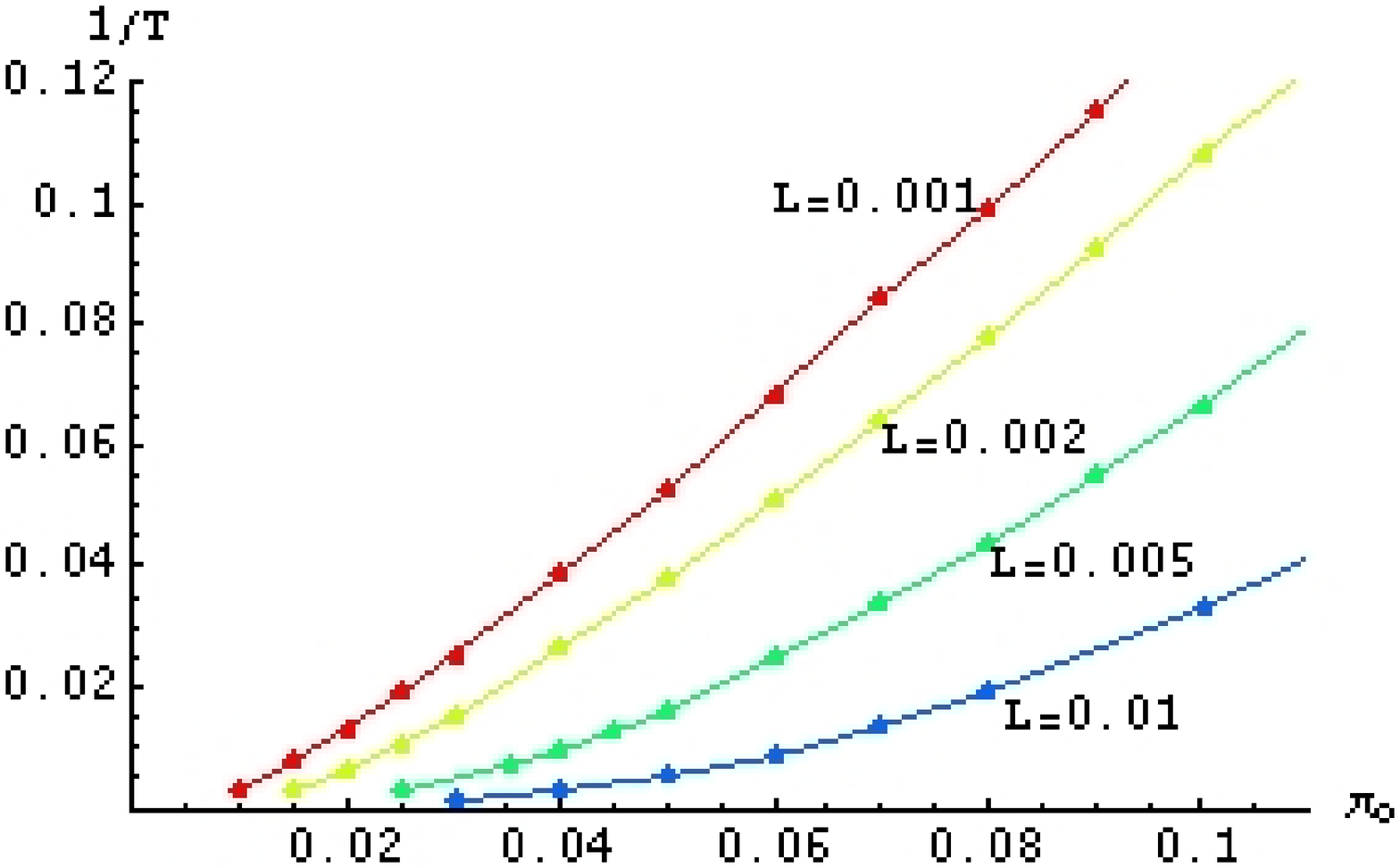,width=3.5in}}
\caption{
\label{fig:singulartime} 
Figure showing the inverse of the time to the `perfect caustic'
singularity, $1/T$, against initial amplitude for various $L$ in
planar symmetry. 
}
\end{figure}

We note that taking sufficiently large $L \sim \sqrt{\pi_0}$ the field
is completely scattered and the behaviour is essentially that of the
linear theory and the non-linear terms never contribute to the
evolution. However this large ghost length is clearly not compatible
with our expectations for structure formation, where we expect $L \ll
|\pi_0| < 1$. For such parameter values the behaviour shown in these
figures \ref{fig:planar1} and \ref{fig:planar2} is generic, and in
figure \ref{fig:singulartime} we plot the (inverse of) the time to the
perfect caustic formation compared to $\pi_0$ for various $L$.  As
expected, this time increases with decreasing amplitude (eventually
crossing over to the smooth linear evolution regime at $L \sim
\sqrt{\pi_0}$ when $T \rightarrow \infty$), or with decreasing
$L$ (so that the radiation from the increasingly perfect central
caustic region is slower). 

Thus for small $L$ a cartoon of the evolution is that the initially
collapsing ghost field causes the higher derivative term to radiate
strongly, but the process of radiation simply refines the collapsing
region into the perfect caustic form where it eventually collapses.

Suitable modification of the conservative equations of motion
(\ref{eq:eom}) introduce the geometric measure factor associated
with axisymmetry. In this case we find the behaviour essentially
analogous to the planar case. Whilst the singular behaviour appears
`weaker', taking longer to reach the singularity for the same
parameters, the singularity does indeed form. However moving to
spherical symmetry we find a change in behaviour.

\begin{figure}[ht]
\centerline{\psfig{file=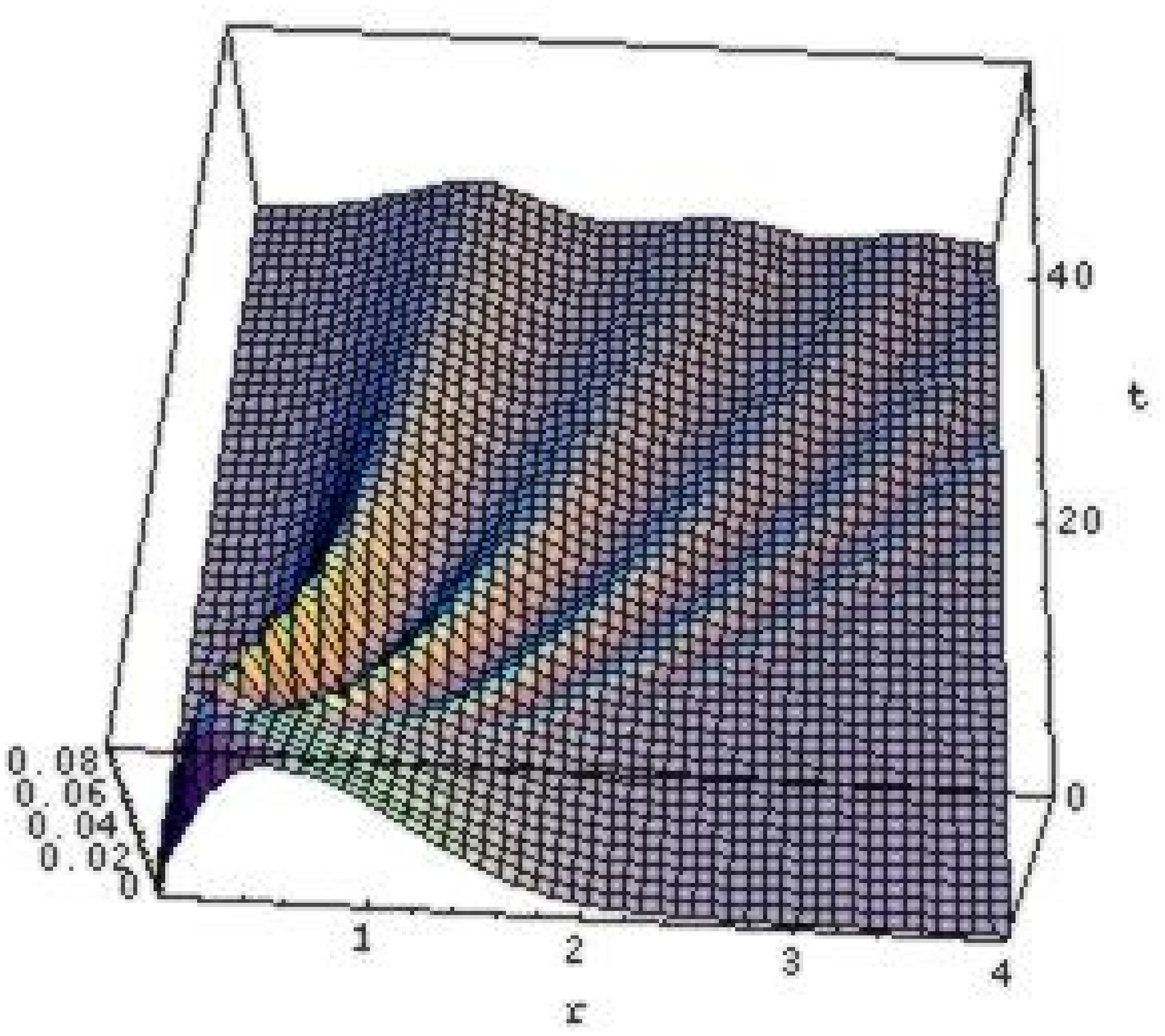,width=3.5in}}
\caption{ \label{fig:sph_nopotl} Plot of $H = \partial_r \pi$ for
spherical symmetry with no gravitational potential. Initial data is
as in the previous planar case, and again the simulation shown has
$\pi_0 = 0.1$ and $L = 0.005$. We see no sign of singularity
formation here, and indeed continued time evolution confirms this. }
\end{figure}

For spherical symmetry we again take the initial data
(\ref{eq:init_data}) and evolve this in the absence of any
gravitational potential. In figure \ref{fig:sph_nopotl} we plot the
evolution of $H$ for $\pi_0 = 0.1$ and $L = 0.005$ (as shown earlier
for the planar case). The behaviour is clearly different with a
totally non-singular evolution for all times, a portion of which is
shown in the figure. Whilst for $L = 0.$ obviously the spherically
symmetric ghost field exhibits the usual caustic singularity, we see
the higher derivative term, aided by the geometric measure factor,
can now radiate sufficient energy to avoid any later energy build
up.

So far we have discussed the case of evolution of a local
perturbation in the ghost field. We have seen that the symmetry of
the initial data has a strong effect on whether the evolution is
singular or not. Physically, however, we are interested more in the
growth of the ghost field in response to a gravitational seed. Thus
we must understand whether singularities in the ghost field form
under these circumstances.

For planar symmetry we may add a gravitational potential to the
system \ref{eq:eom} by simply introducing a potential $\Phi$ to
modify $\Sigma$ as,
\begin{equation}
\dot{H} = \partial_r \left[  \Sigma + \frac{1}{2} H^2 + \Phi \right]
\end{equation}
We then use initial data for the evolution where both $H$ and $\Sigma$
are zero, and allow the gravitational potential to seed the growth
in the ghost field. We find that for the weak potentials expected in
our physical context, in the planar case this does not influence the
behaviour appreciably. The gravitational source simply serves to
produce a local non-zero configuration of the ghost field, which,
once formed, then undergoes the singular evolutions described
earlier, and is then largely independent of the presence of the
potential. The same is true for axisymmetry.

\begin{figure}[ht]
\centerline{\psfig{file=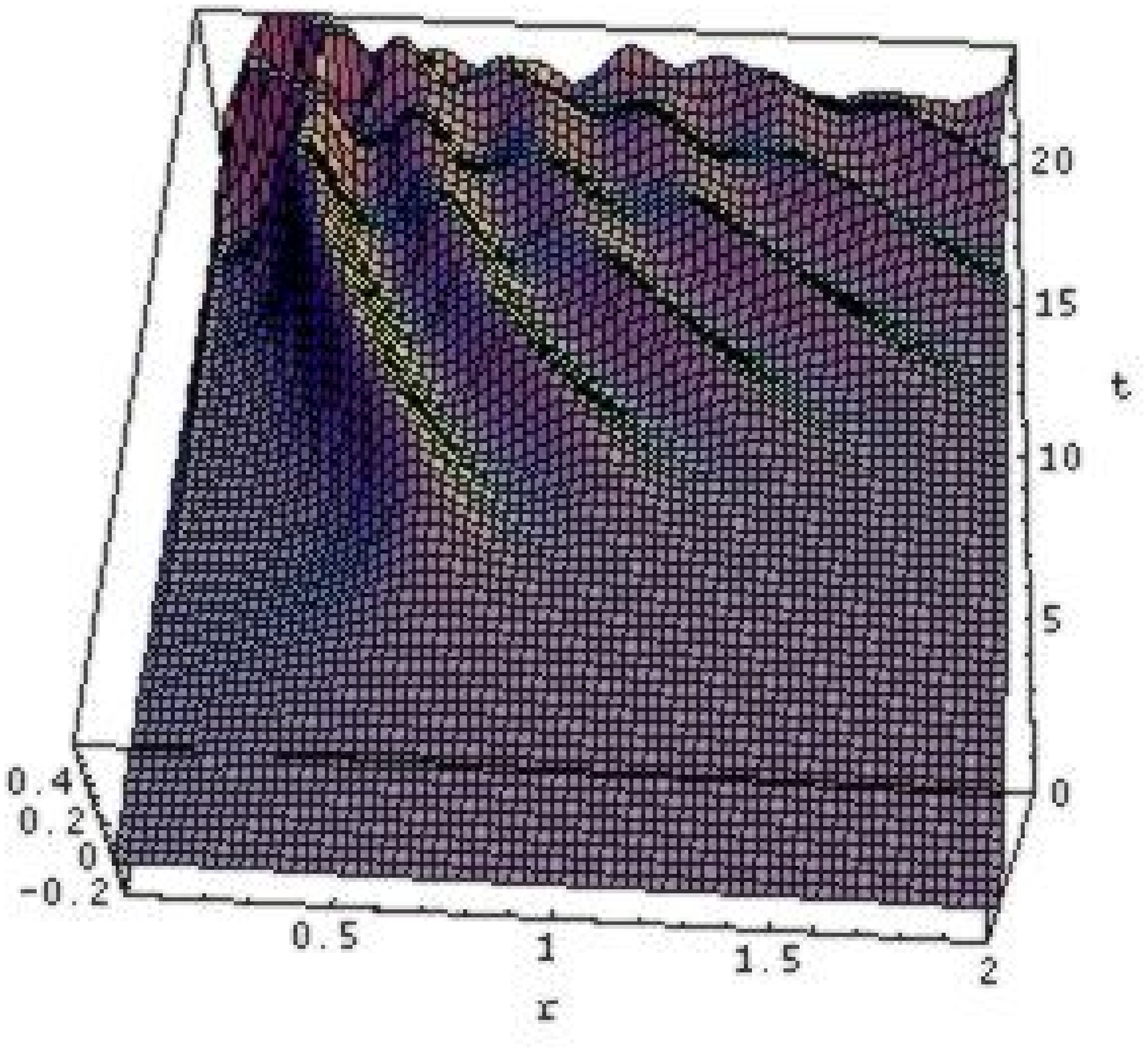,width=3.5in}
\psfig{file=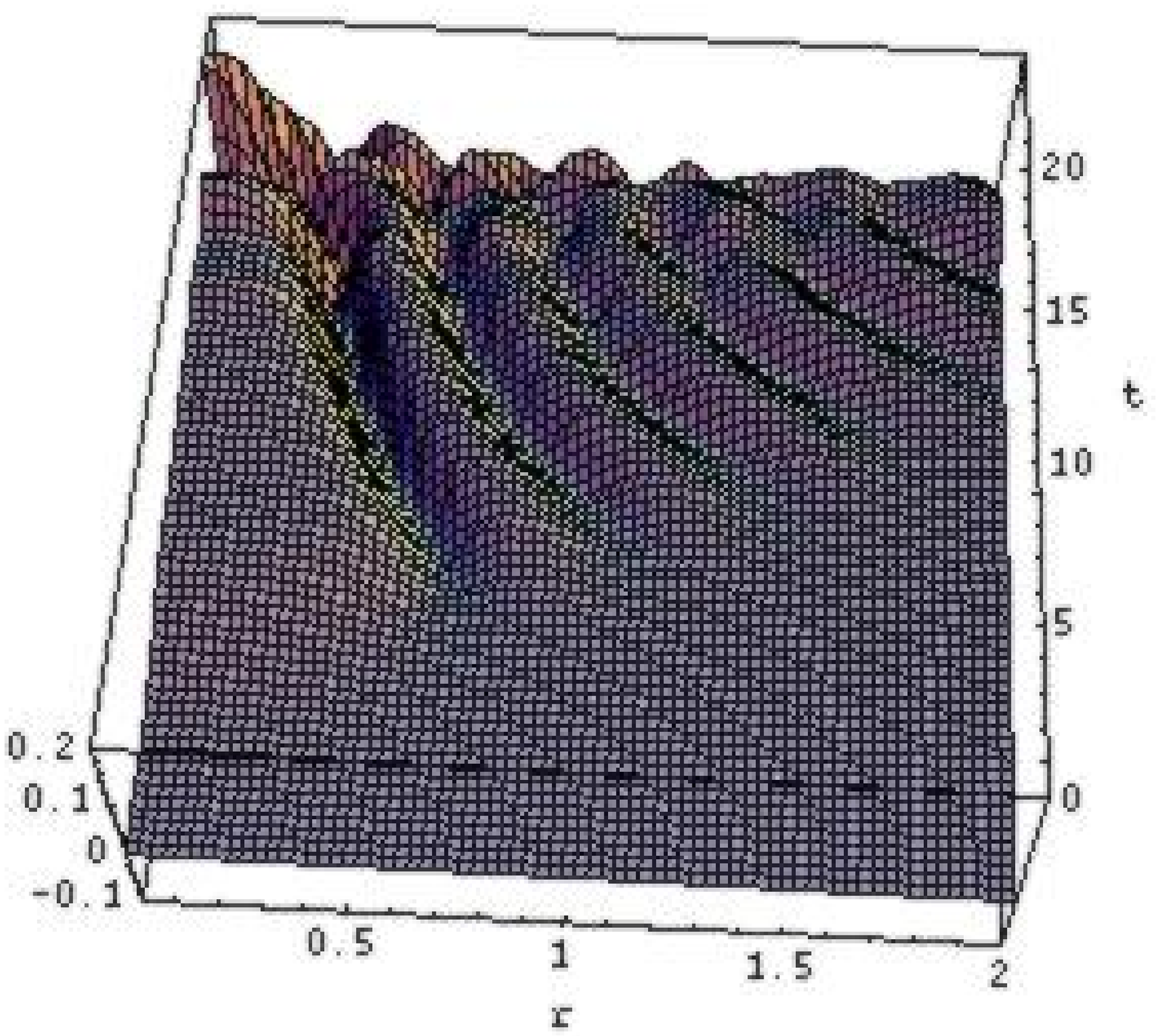,width=3.5in}}
\caption{ \label{fig:sph_potl} Plot of $H = \partial_r \pi$ and $\Sigma$
for spherical symmetry with a gravitational potential. We see that a
singularity does eventually form, whereas earlier we saw that with
no potential, smooth localized initial data evolves in a
non-singular manner for spherical symmetry. }
\end{figure}

The interesting case is clearly spherical symmetry, as without a
potential we find a non-singular evolution for smooth localized
initial data. We take a potential such that,
\begin{equation}
\nabla^2 \Phi = \Phi_0 e^{-r}
\end{equation}
In analogy with our earlier simulations $\Phi_0$ now acts like
$\pi_0$ and we have taken units so the physical size of the source
is of order unity. Then in figure \ref{fig:sph_potl} we show a
typical evolution of $H$ and $\Sigma$ for parameters $\Phi_0 = 0.2$ with
$L = 0.02$.

We now see a complicated evolution where waves are radiated early on
as the higher derivative term becomes important near the symmetry
axis, but later these waves, consisting of positive and negative $\Sigma$
regions, are drawn grudgingly into the centre, eventually forming a
singularity. This contrasts with the case without potential where
such waves form, but are quickly radiated to infinity.

Thus the less singular behaviour present in spherical symmetry
appears to be overcome once a gravitational potential is introduced
and we conclude that, with the choice of higher derivative term in
(\ref{eq:eom}), the evolution of the ghost field in response to
gravitational sources for all 3 reduced symmetries is singular.

\subsection{Three dimensional evolution}

We now test our claim that evolution is singular for gravitational
seeds. In particular it is not obvious which symmetry will dominate
the singularity evolution. For a point seed one might assume
spherical symmetry to be the most relevant. However we have seen it
is in the planar case that the singularity formation is quickest.

We evolve the $\pi$ field itself in three dimensional Cartesian
coordinates using a simple explicit second order method. Clearly
here resolution becomes an important issue and we are less able to
separate the 2 length scales in the problem. However even with very
modest resources and resolutions ($\simeq 150^3$) we can achieve
useful results.

Again we use initial conditions where $\pi$ and $\Sigma$ are zero. We
seed the dynamics of the ghost field by taking a Gaussian potential,
\begin{equation}
\Phi = - \Phi_0 e^{- r^2}
\end{equation}
where now $r^2 = (x - x_0(t))^2 + (y - y_0(t))^2 + z^2$ and,
\begin{equation}
x_0(t) = 0.7 \cos{t} \qquad y_0(t) = 0.7 \sin{t}
\end{equation}
so our gravitational seed executes a circular motion in the $z=0$
plane. Hence our system now has no continuous symmetries although
there is a preferred plane, and also the geometry of the source is
locally spherical. Therefore we may gain information about the local
geometry of any singularities that form.

\begin{figure}[ht]
\centerline{\psfig{file=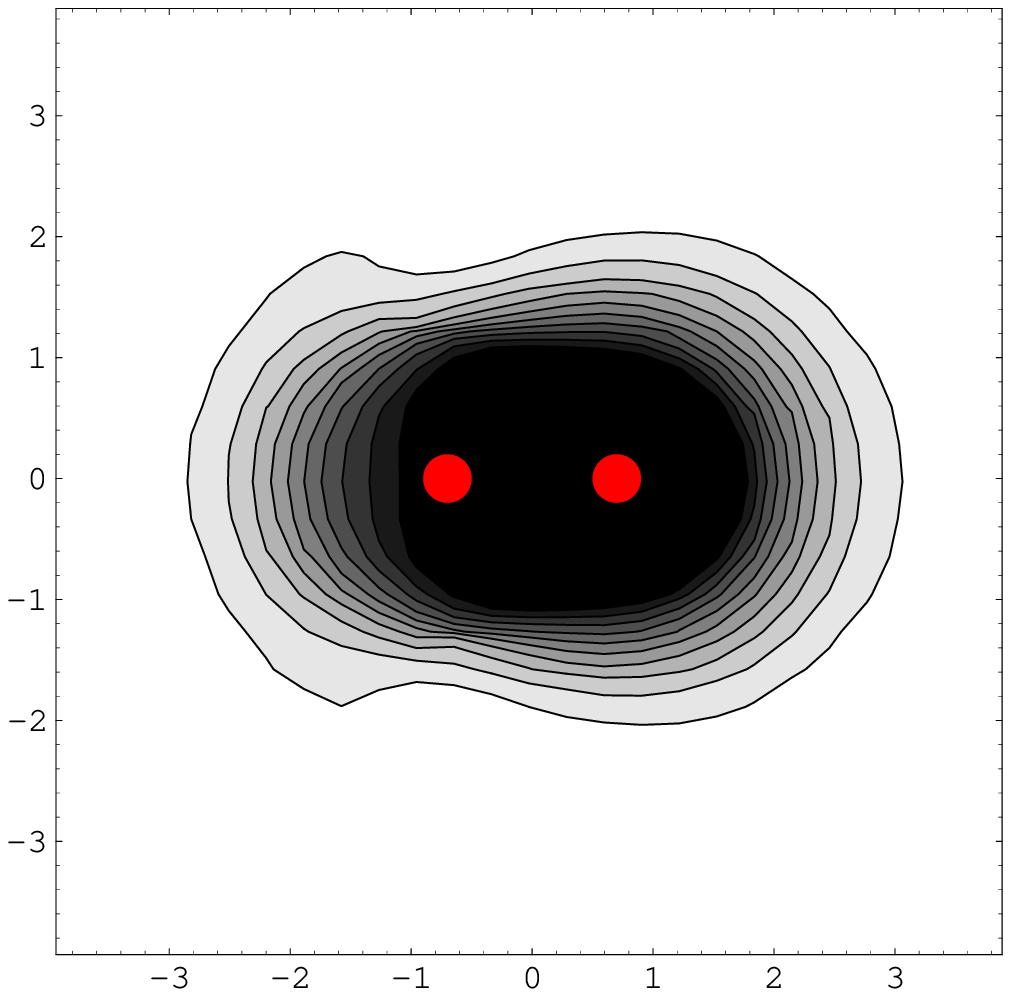,width=3.5in}
\psfig{file=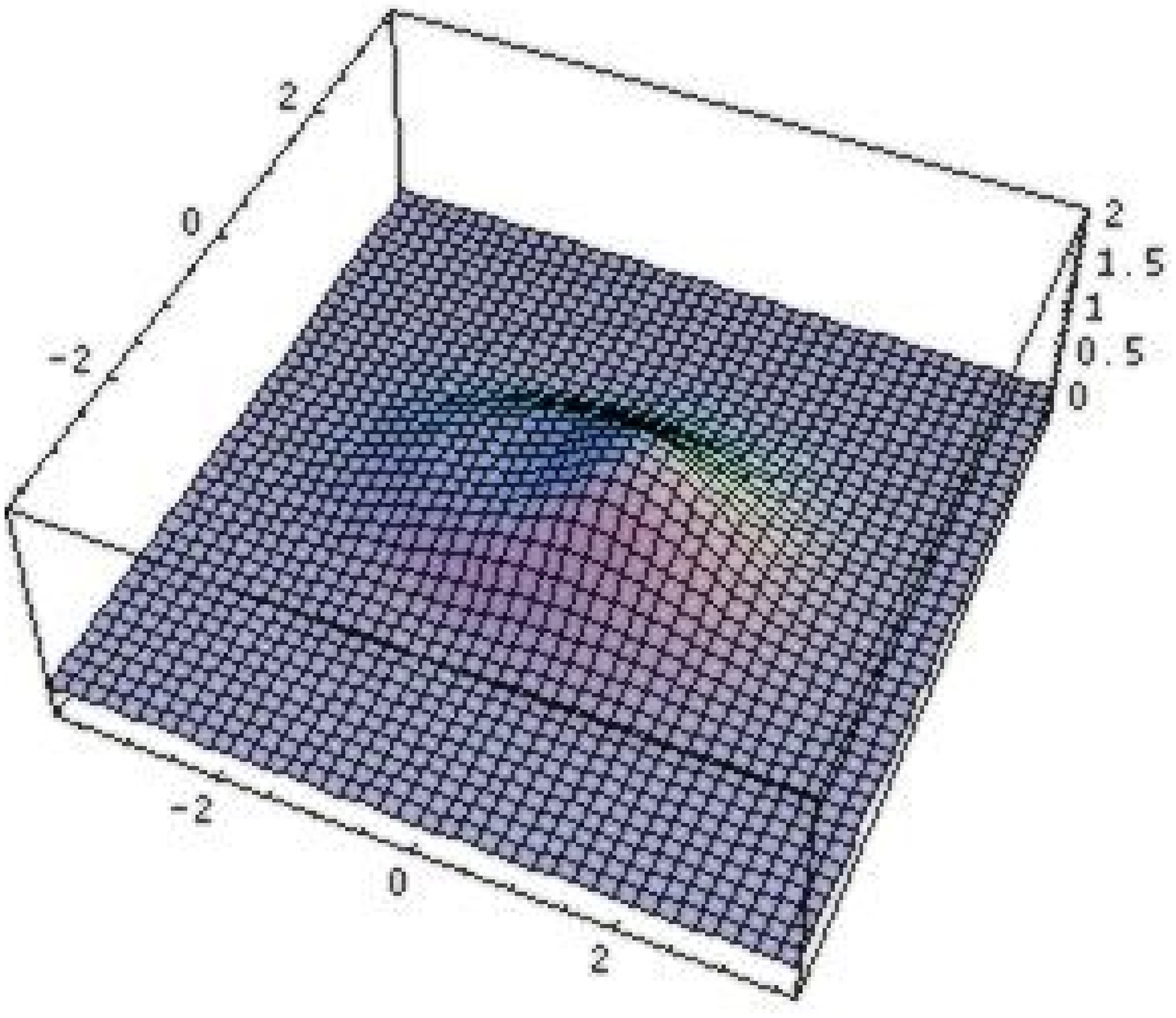,width=3.5in}} \caption{
\label{fig:3d_a} Figure showing $\pi$ in the $x=0$ plane at time $t
= 2.0$ for a full 3-d evolution. The ghost field is sourced by a
slowly moving gravitational potential, orbiting in the $z=0$ plane.
The intersection points of the orbit with the slice shown are
indicated by red dots in the contour plot. The peak in the field
reflects the recent passage of the source through the plane.  }
\end{figure}

\begin{figure}[ht]
\centerline{\psfig{file=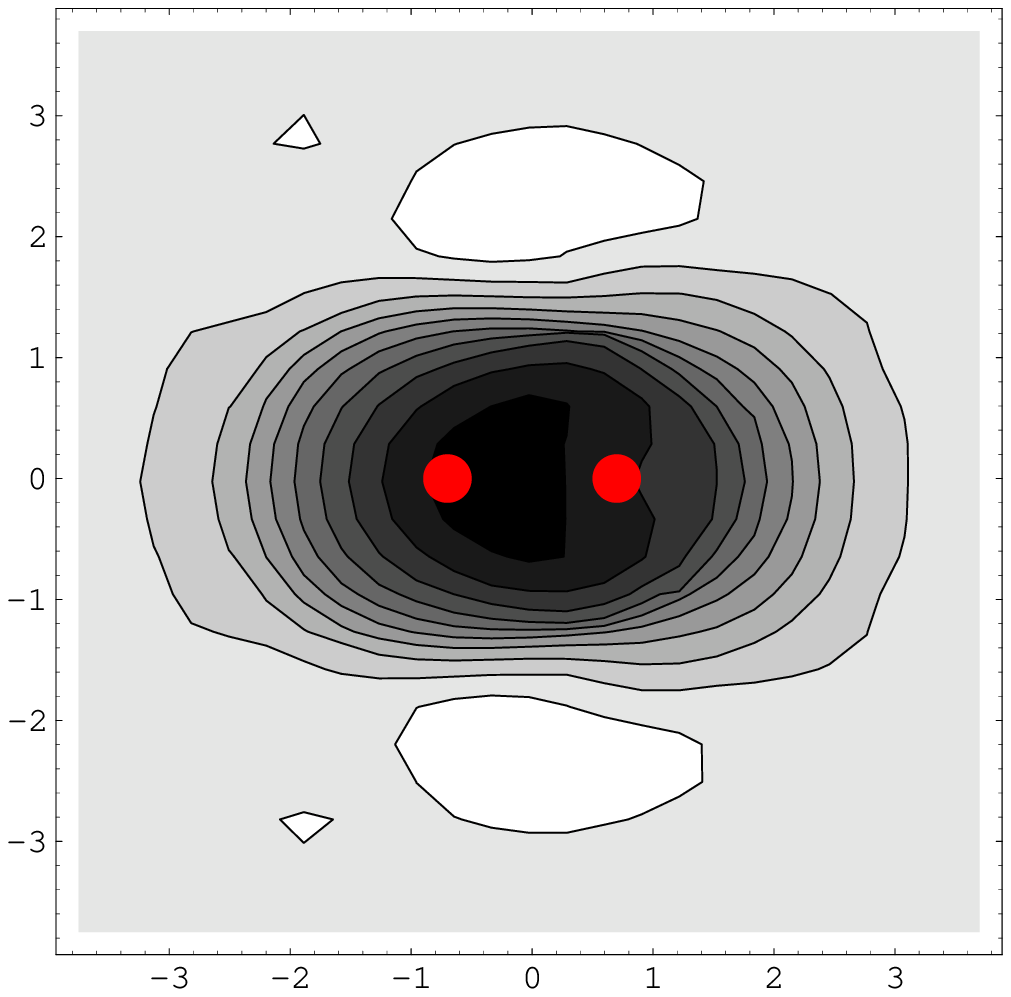,width=3.5in}
\psfig{file=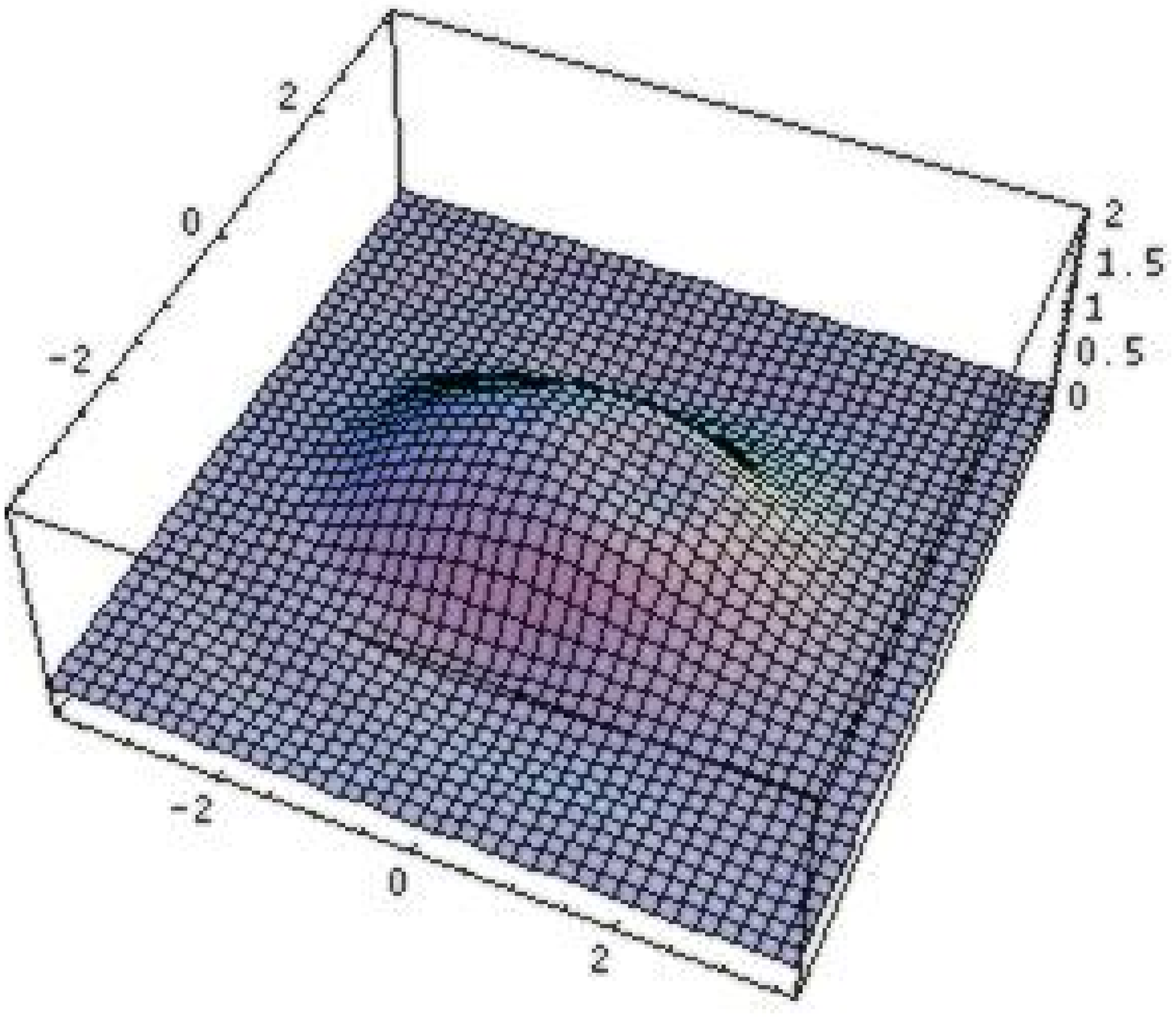,width=3.5in}} \caption{
\label{fig:3d_b} Figure showing an intermediate time slice of $\pi$
at $t = 4.0$ for the same 3-d evolution as in the previous figure.
}
\end{figure}

\begin{figure}[ht]
\centerline{\psfig{file=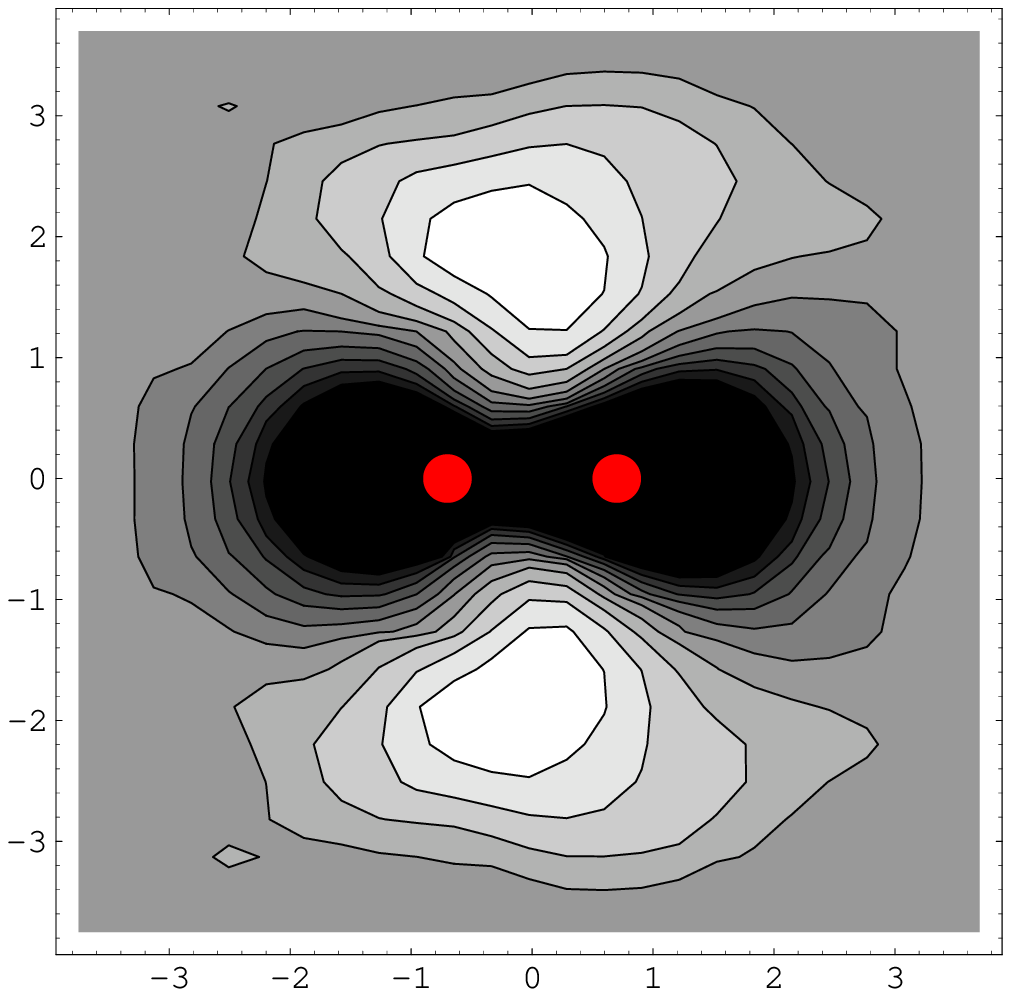,width=3.5in}
\psfig{file=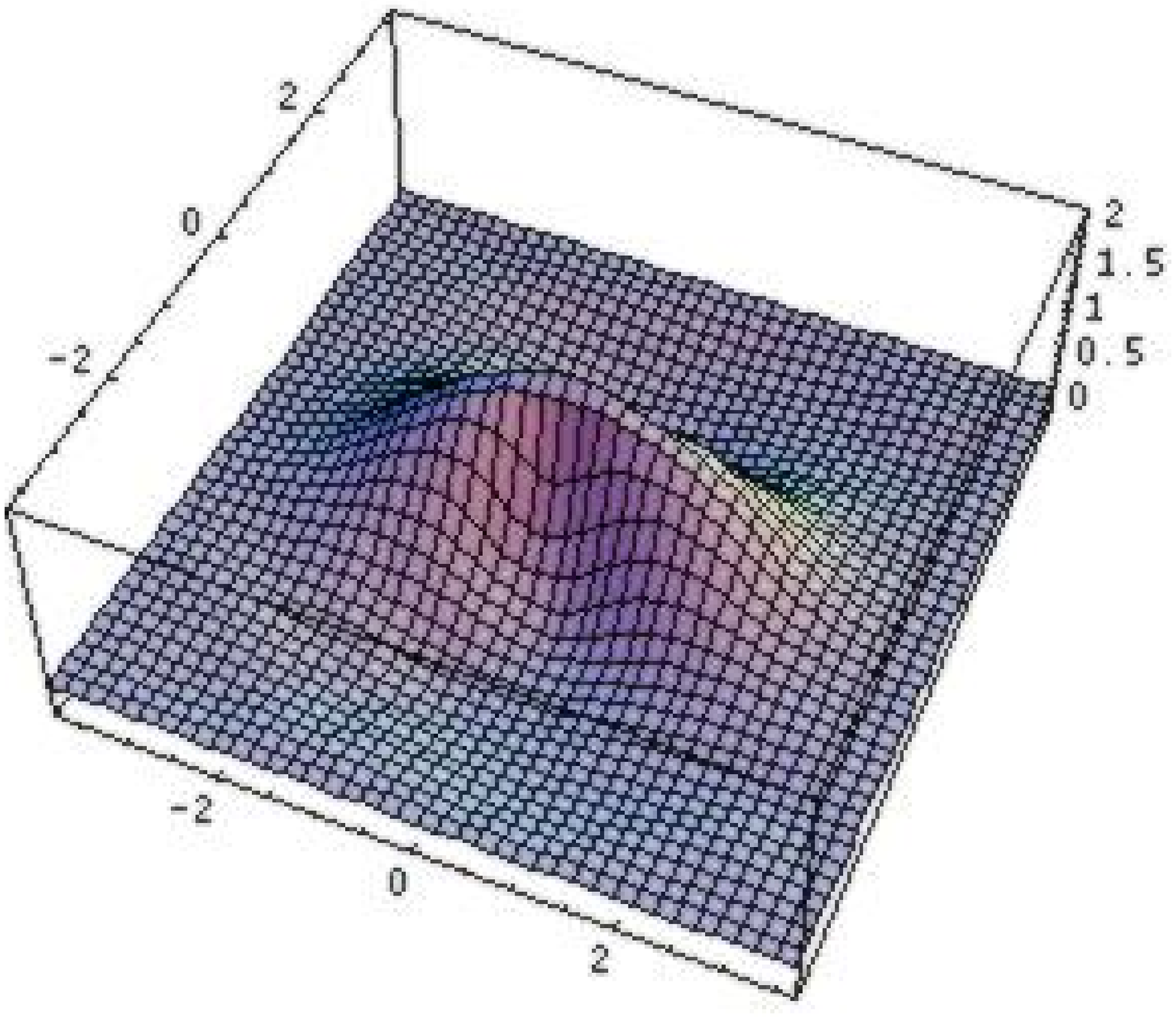,width=3.5in}} \caption{
\label{fig:3d_c} Continuing from the previous figures this plot
shows the time slice $t = 6.1$, just before the field becomes
singular. Note the geometry of the almost singular field appears
planar, and is roughly confined in the disc bounded by the source
motion.  }
\end{figure}

Figures \ref{fig:3d_a}, \ref{fig:3d_b} and \ref{fig:3d_c}, show
$\pi$ in the $x=0$ plane at various times for a typical evolution
with parameters $\Phi_0 = 2.$ and $L = 0.2$ (note the physical
scales are not so well separated as in the one dimensional
examples). The $y=0$ slices look very similar.  We see that, as
indicated from our study of the reduced dimension examples above, a
singularity does indeed form, the last time frame being immediately
before the configuration becomes singular. For the same static
source, so $x_0(t) = y_0(t) = 0$, the singularity occurs at $t
\simeq 3.1$, but for the moving seed described here, it occurs at $t
\simeq 6.2$, and the source executes approximately one revolution.

We see from the figures, and also the complete 3-d data, that the
singularity develops in the $z = 0$ plane in which the source moves,
and appears to have a planar form, filling the disc bounding the
motion of the source. This is despite the fact that the source only
completes one `orbit' and thus is sufficiently slow moving that this
disc symmetry is not obviously imposed. Thus we take this as evidence
confirming our naive expectation that the local geometry of the
singularity will be the symmetry where singularity formation is
strongest, namely planar.

\subsection{Interpretation of numerics}

In subsection~\ref{subsec:negativeenergy} we have shown that a
negative $\Sigma$ region does not grow but shrinks asymptotically. To
be precise, the acceleration of the boundary of a negative energy
region is towards the negative region itself. It has also been argued
in subsection~\ref{subsec:causticsoln} that the perfect caustic is an
attractor. With numerical results at hand, it is easy to confirm
them. In this subsection, for simplicity we illustrate the
confirmation in the planar case only, while the same qualitative
features are seen also in the spherical case with the potential.

It is certainly worth while asking how caustics will form during
structure formation. Here, we define a caustic in the broad sense as a
region where $\Sigma$ becomes $O(1)$. Our understanding of caustics is
still primitive and this question is obviously beyond the scope of
this paper, but we can at least say that a caustic region does not
swallow the whole space. The essential reason for this is the
existence of a conserved charge $\int \Sigma dv$ resulting from the
shift symmetry: if $|\Sigma|$ grows then the volume shrinks. In
particular, we have shown and numerically confirmed that a negative
$\Sigma$ region shrinks. In this sense caustics in the ghost
condensate are somewhat similar to shockwaves in hydrodynamics. For a
shockwave in hydrodynamics, we do not need to specify microscopic
physics, such as atomic physics, to describe its dynamics and
influence to macroscopic physics outside. Therefore, it is expected
that we should be able to predict what happens outside the caustic
region without too many assumptions about a UV completion in the
caustics. In the next subsection we shall discuss possible UV
completions.

Based on our numerical result, we expect that a planar caustic, being
pretty much like a pancake or a disk, forms during structure
formation. Indeed, the 3-d simulation showed that the initial stage
looks spherical but deviations from spherical symmetry grow and the
system evolves towards a planar shape.

As future work, we should ask at what scale caustics forms first in
realistic situations. In particular, it is interesting to ask whether
(i) a large caustic forms first and cascades to smaller ones via
fragmentation or (ii) many small caustics form and buildup to large
caustics. In either case, we might be tempted to speculate that the
distribution of caustics should somehow trace the matter distribution
since the formation of caustics is enhanced by the gravitational
potential due to matter. Clearly, we need more detailed analysis to
see if this is the case or not. For this purpose we of course need to
specify the initial condition of $\pi$ and the matter distribution,
which determines the external gravitational potential. To be specific,
let us suppose ghost inflation~\cite{Arkani-Hamed:2003uz} happened at
$H\simeq 10^{-4}M$. This is sufficient to explain the amplitude of
primordial density perturbations $\delta\rho/\rho\simeq 10^{-5}$,
which determines the initial conditions for matter, or equivalently
the gravitational potential. We also have primordial perturbations of
$\pi$. Thus, the problem to be solved is well-defined at least in
principle. The nonlinear dynamics of the matter-$\pi$ coupled system
with this initial condition certainly deserves further investigation.



\subsection{Possible UV completion/extension: gauged ghost and/or new
  terms}

We have seen that the nonlinear dynamics leads to formation of
caustics. For the perfect caustics, the $\nabla^4\pi$ term in the
equation of motion vanishes and, thus, does not act as a leading
spatial derivative term. This is the very reason why the perfect
caustic does not bounce. At the same time, this means that the spatial
derivative expansion in this background starts differently from that
in the trivial $\pi=0$ background and, thus, new terms should be
included in the leading Lagrangian for the perfect caustic
background. It is also possible to consider new degrees of freedom
which become relevant as the system approaches the caustic
singularity.

With this in mind, for practical purposes (such as numerical studies of
the ghost condensate as a dark matter candidate) it is probably useful
to have simple models of new terms and/or new degrees of freedom which
manifestly cutoff the singular behavior. Implementations adopted in this
subsection are not particularly based on a consistent derivative
expansion nor a more fundamental theory. Nonetheless, models here may be
useful for practical purposes and, at least, show the existence of
regular systems without caustics.

First, we consider the introduction of new degrees of freedom by 
turning on the gauge coupling for the gauged condensation proposed in
ref.~\cite{Cheng:2006us}. (Inclusion of new terms will be considered
later in this section.) The gauged ghost condensation has new degrees of
freedom, but they decouple from $\pi$ in the limit of zero gauge
coupling. Hence, by promoting the gauge coupling constant to a function
of $\pi$'s derivatives it is possible to make the new degree relevant
only near the singularity.

The leading Lagrangian for the gauged ghost condensation is
obtained by gauging the shift symmetry and is of the following
form. 
%
\begin{equation}
 \scr{L} = \frac{M^2}{2}\left(\scr{A}_0 - M\Phi -\frac{1}{2M} 
\vec{\scr{A}}^2\right)^2
- \frac{1}{4g^2} F_{\mu\nu} F^{\mu\nu} - \frac{\alpha}{2}
(\grad \cdot \vec{\scr{A}})^2.
\end{equation}
In the $g^2\to 0$ limit, $F_{\mu\nu}$ goes to zero and, thus,
$\scr{A}_{\mu}$ is written as a derivative of a scalar:
$\scr{A}_{\mu}=\partial_{\mu}\pi$. Therefore, in this limit the
original (ungauged) ghost condensation is recovered. In
ref.~\cite{Cheng:2006us} the nonlinear dynamics of the gauged ghost
condensation was analyzed. It was shown that gauging the ghost
condensation resolves the caustics for the spherically symmetric case
with an external gravitational force. On the other hand, the caustics
in the perfectly planar symmetric case remains unresolved. It was
argued that small fluctuations on top of the perfectly planar
symmetric layer should grow and lead to fragmentation of the layer
into small pieces and caustics with codimension more than one should
not form after that. Thus, it was argued that the gauged ghost
condensation does not lead to caustics except for extremely fine-tuned
situations. In the following we shall implement the gauged ghost
condensation as a possible UV completion. As a very simple
implementation, the gauge coupling is turned on when and only when
$\vec{\scr{A}}^2$ is large. We shall see that this indeed cuts off the
caustics in the spherically symmetric case with external gravitational
force. On the other hand, we do not expect to cutoff the caustics in
the perfectly planar symmetric case since the gauged ghost
condensation itself does not bounce the perfectly planar symmetric
caustics. However, as argued in ref.~\cite{Cheng:2006us} and stated
above, a planar symmetric caustics with perturbation is expected to
fragment into small pieces and to bounce eventually.

With the spherical symmetry, the equation of motion is reduced to 
%
\begin{eqnarray}
 \dot{\Sigma} & = & \frac{1}{r^2}
  \left\{ r^2\left(\Sigma\xi - b\left[\frac{1}{r^2}(r^2\xi)'\right]'
             \right)\right\}', \nonumber\\
 \dot{\xi} & = & G^2b f + \Sigma' + \xi\xi' + \Phi', \nonumber\\
 \dot{f} & = & -\frac{1}{b}
  \left\{\Sigma\xi - b\left[\frac{1}{r^2}(r^2\xi)'\right]'
                   \right\},
\end{eqnarray}
where 
%
\begin{equation}
 \Sigma \equiv \frac{1}{2}(X-1), \quad
 \xi \equiv D_r\phi, \quad
 f \equiv \frac{1}{G^2b}\frac{F_{tx}}{ML_{phys}},
\end{equation}
%
\begin{equation}
 b = \frac{\alpha}{(ML_{phys})^2}, \quad
  G = gML_{phys}.
\end{equation}
We can eliminate $\Sigma$ from the set of equations essentially because the
electric field $f$ is determined solely by the charge. Actually, since
%
\begin{equation}
 \left[\frac{b}{r^s}(r^sf)'+\Sigma\right]^{\cdot} = 0,
\end{equation}
the combination in the squared bracket is independent of $t$. Since the
expansion of the universe in the early universe dilutes $\Sigma$ and $f$ to
zero with an extremely high precision, we set
%
\begin{equation}
 \Sigma|_{t=0} = f|_{t=0} = 0.
\end{equation}
Hence, we obtain
%
\begin{equation}
 \Sigma = -\frac{b}{r^2}(r^2f)',
\end{equation}
and $\Sigma$ can be eliminated from the set of equations. The result is 
%
\begin{eqnarray}
 \dot{\xi} & = & G^2bf - b\left[\frac{1}{r^2}(r^2f)'\right]'
  + \xi\xi' + \Phi', \nonumber\\
 \dot{f} & = & \frac{1}{r^2}(r^2f)'\xi
  + \left[\frac{1}{r^2}(r^2\xi)'\right]'.
\end{eqnarray}

\begin{figure}[ht]
\centerline{\psfig{file=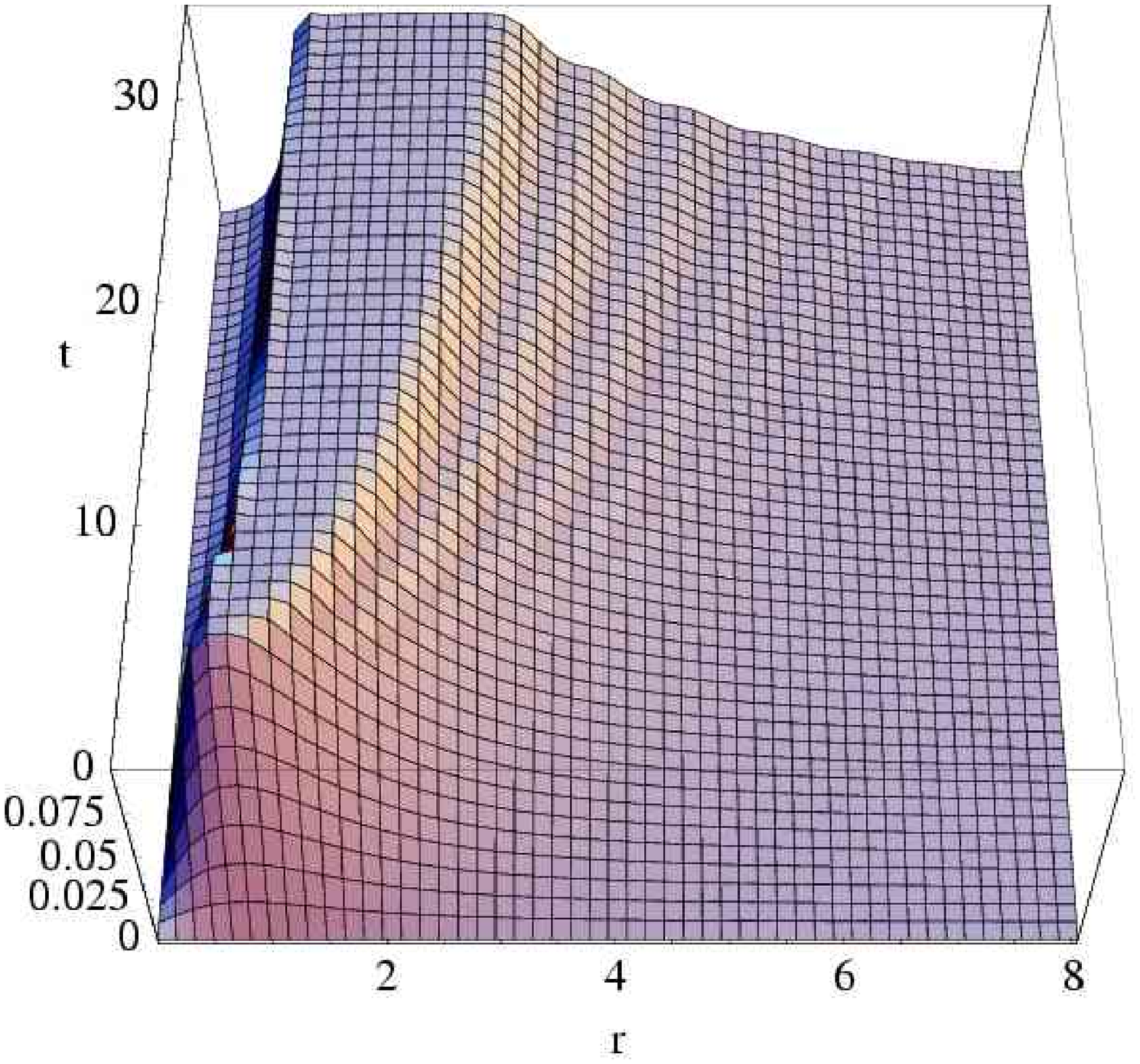,width=3.5in}
\psfig{file=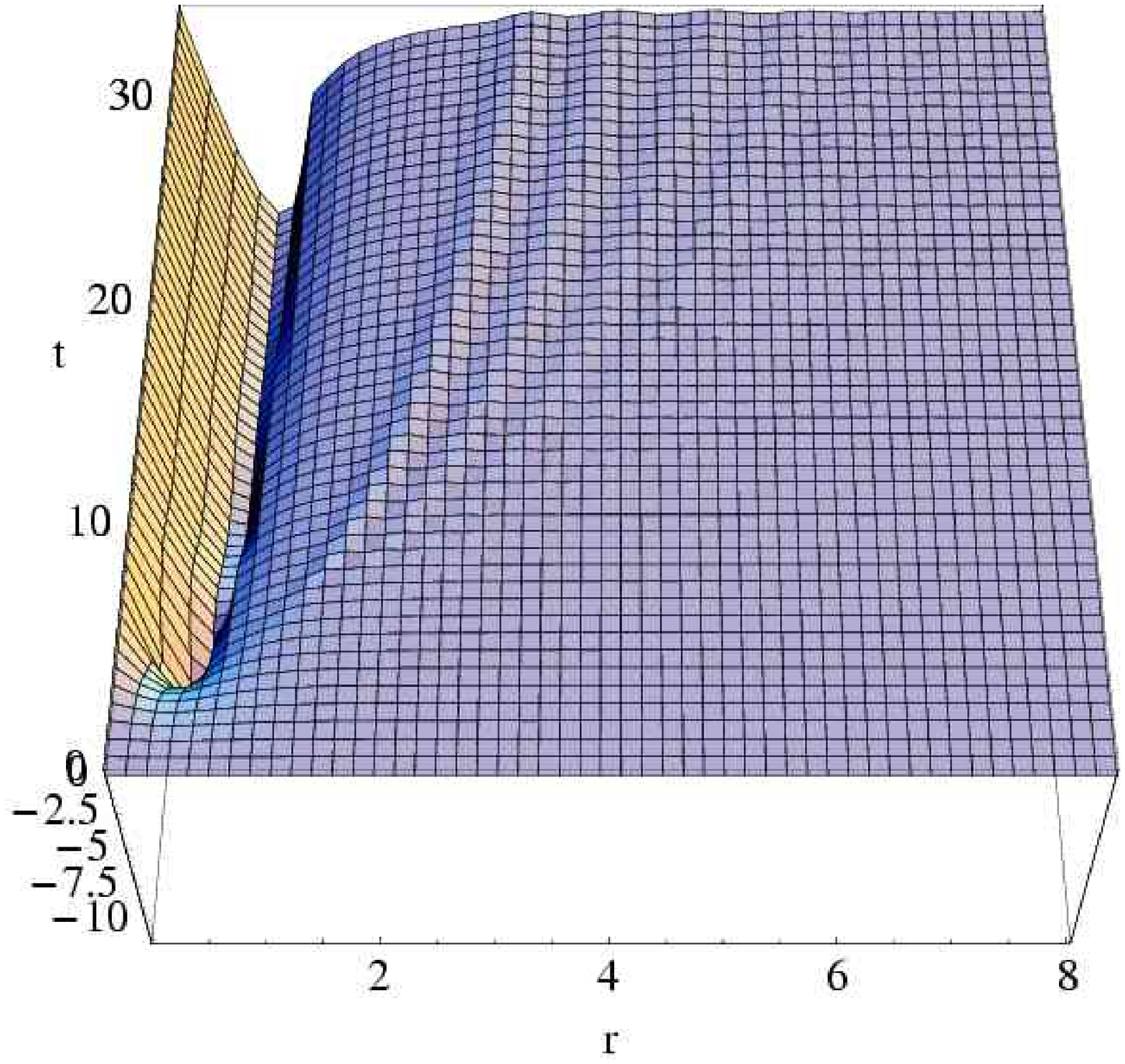,width=3.5in}}
\caption{ \label{fig:gauged}
 The result of numerical simulation with $\alpha=2.0\times 10^{-4}$,
 $G_0^2=1.0\times 10^3$, $\xi_0=1.0\times 10^{-1}$ and 
 $\delta=5.0\times 10^{-4}$. The left figure shows $\xi$. The right
 figure shows $f$. It is easy to see in the left figure that $|\xi|$
 larger than $\xi_0$ is cut off by the gauge coupling. It is also
 easy to see in the right figure that the implosive accumulation of
 incoming waves is supported by the repulsive electric force ($f<0$)
 near the center. 
}
\end{figure}

\begin{figure}[ht]
\centerline{\psfig{file=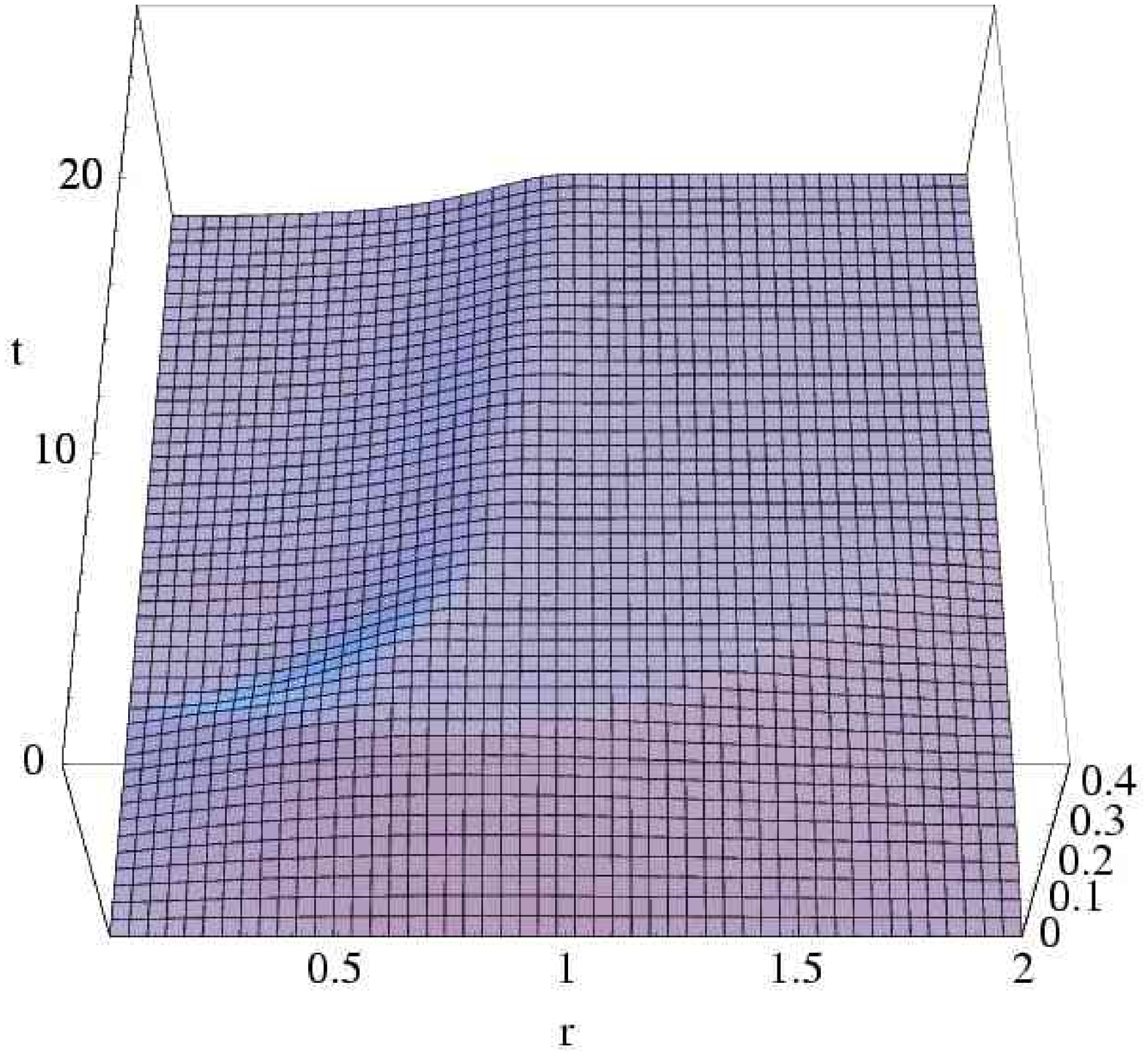,width=3.5in}
\psfig{file=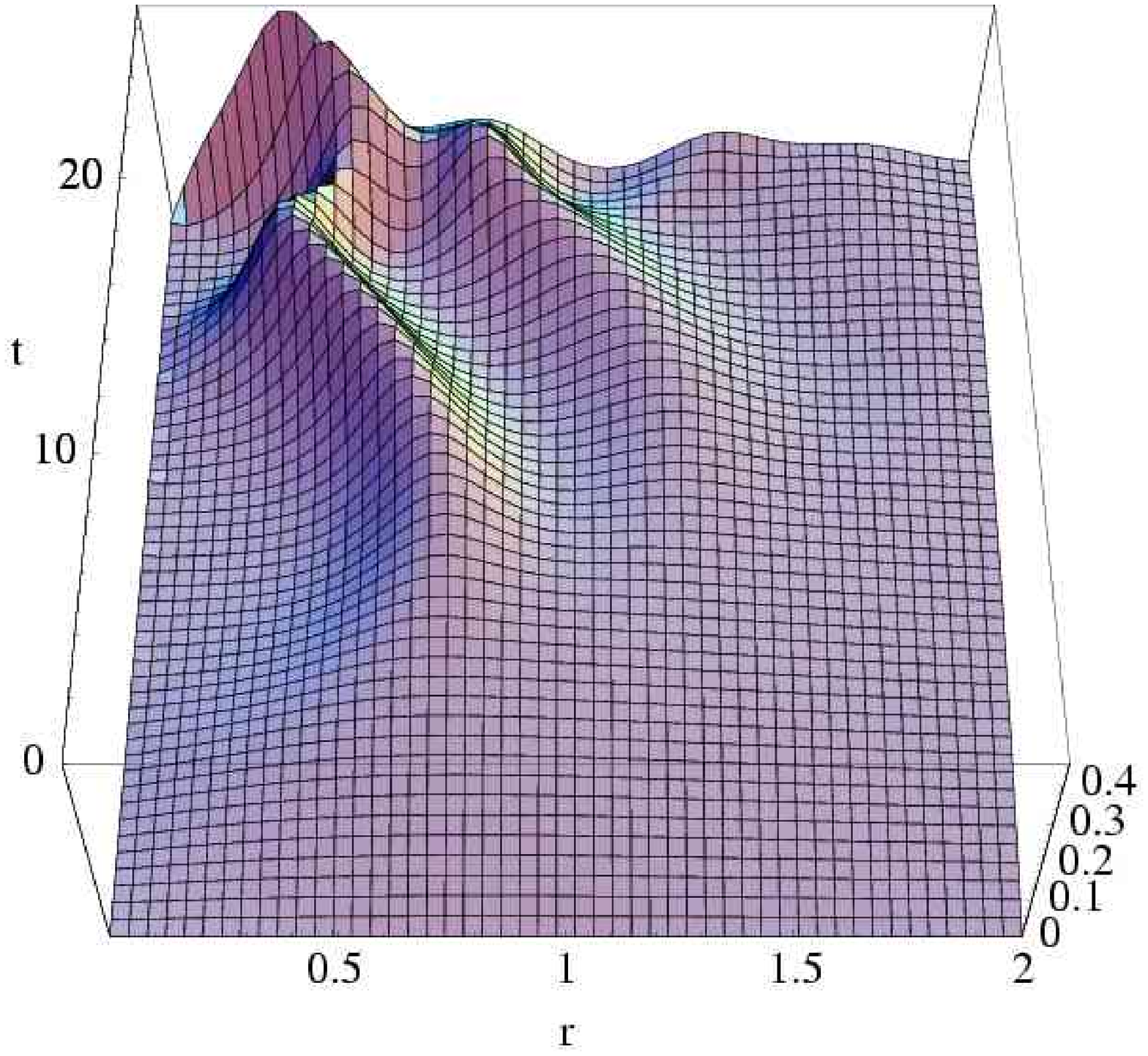,width=3.5in}}
\caption{ \label{fig:with-without-g}
 Comparison of the regular result for $G_0^2=1.0\times 10^3$ with the
 singular result for $G_0^2=0$. The left  figure shows a closer look at
 the regular behavior of $\xi$ shown in Fig.~\ref{fig:gauged}. The
 right figure shows the singular behavior of $\xi$ for $G_0^2=0$. The
 value of $\alpha$ and the external gravitational potential $\Phi$ are
 the same as before. Note that the scale of hight is different for these
 two figures.
}
\end{figure}

We have described the gauged ghost condensation proposed in
ref.~\cite{Cheng:2006us}. Now we promote $G^2$ from a constant to a function of 
$\xi^2$ so that the gauge coupling is turned on when and only when the
gradient becomes sufficiently large. For simplicity we choose 
%
\begin{equation}
 G^2 = \frac{G_0^2}{2}
  \left[ 1 + \tanh\left(\frac{\xi^2-\xi_0^2}{\delta^2}\right)\right],
\end{equation}
where $G_0$, $\xi_0$ and $\delta$ are constants.
Numerical simulation with $\alpha=2.0\times 10^{-4}$, 
$G_0^2=1.0\times 10^3$, $\xi_0=1.0\times 10^{-1}$ and 
$\delta=5.0\times 10^{-4}$ is shown in
Fig.~\ref{fig:gauged}. Fig.~\ref{fig:with-without-g} compares this case 
with the singular result for the ungauged ghost condensation,
i.e. $G_0=0$.

Now, let us turn our attention to possible new terms which manifestly 
cutoff the singular behavior. For this purpose it is useful to consider
the nonlinear Hamiltonian of the system and to seek new terms which have
positive contribution to the Hamiltonian. For simplicity we restrict our
consideration to terms depending on $z\equiv (\nabla\pi)^2$ only.

To the leading Lagrangian $L_0=(X-1)^2/2-\alpha(\nabla^2\pi)^2$, let us
add a new term of the form $-F(z)$:
%
\begin{equation}
 L = \frac{1}{2}(X-1)^2  -\alpha(\nabla^2\pi)^2
  - F(z), \quad
  X-1 = 2\dot{\pi} - z, \quad z = (\nabla\pi)^2.
\end{equation}
The Hamiltonian, or the generator of the unbroken diagonal part of (time
translation symmetry)$\times $(the shift symmetry), is 
%
\begin{equation}
 H = 2\dot{\pi}^2 - \frac{1}{2}z^2 + \alpha(\nabla^2\pi)^2 + F(z). 
\end{equation}
The second term $-z^2/2$ has negative sign and, thus, the Hamiltonian
would be unbounded from below without new terms. Actually, this is the 
origin of the nonlinear instability. Hence, in order to stabilize the
system we would like to make the Hamiltonian bounded from below by
choosing the new term appropriately. This is indeed the case if $F(z)$
increases faster than $z^2/2$ for large $z$. On the other hand, $F(z)$
must not change physics below cutoff $M$ and, thus, must be turned off
for small $z$. In particular, it must at least satisfy $F'(0)=0$.

\begin{figure}[ht]
\centerline{\psfig{file=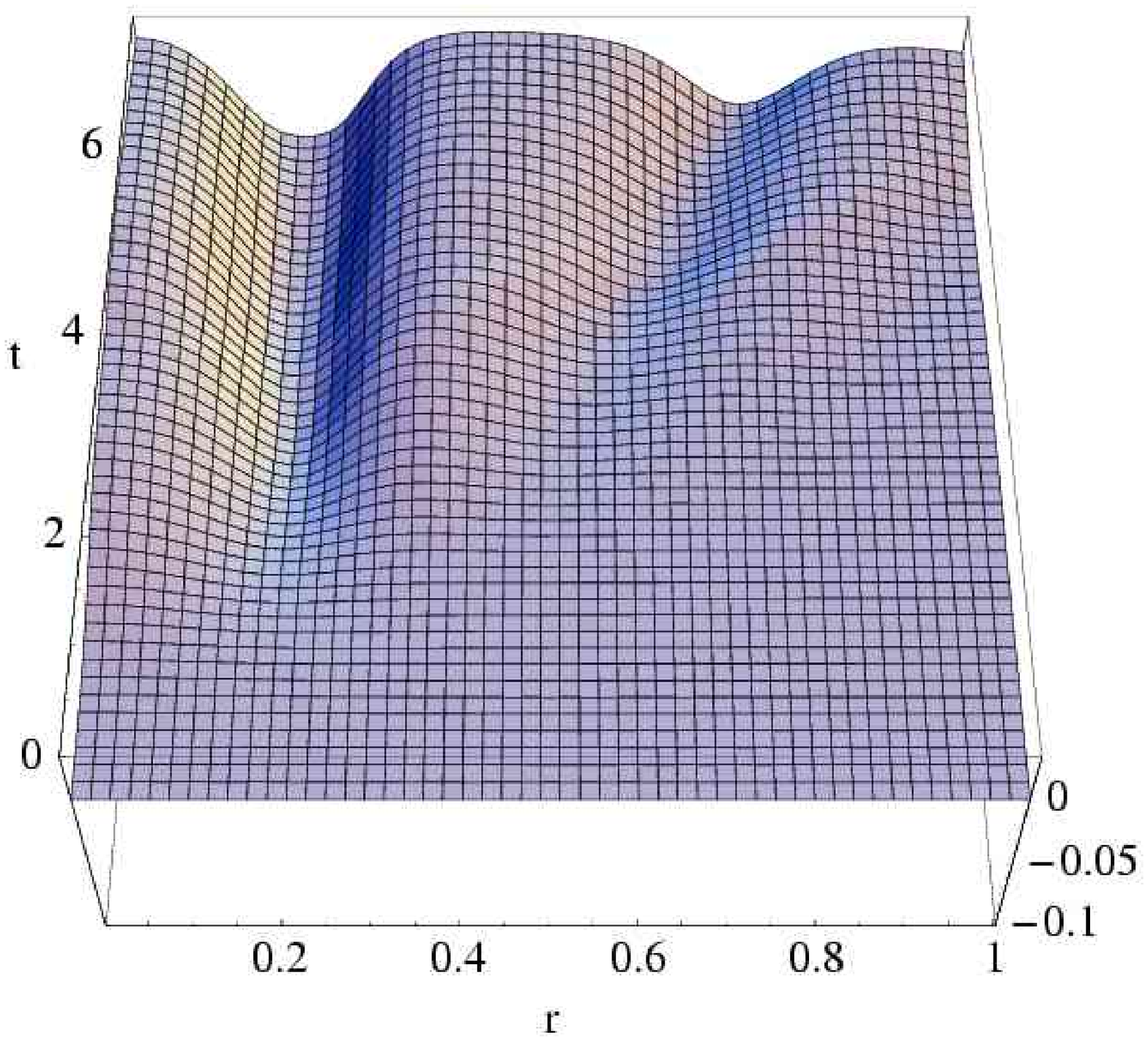,width=3.5in}
\psfig{file=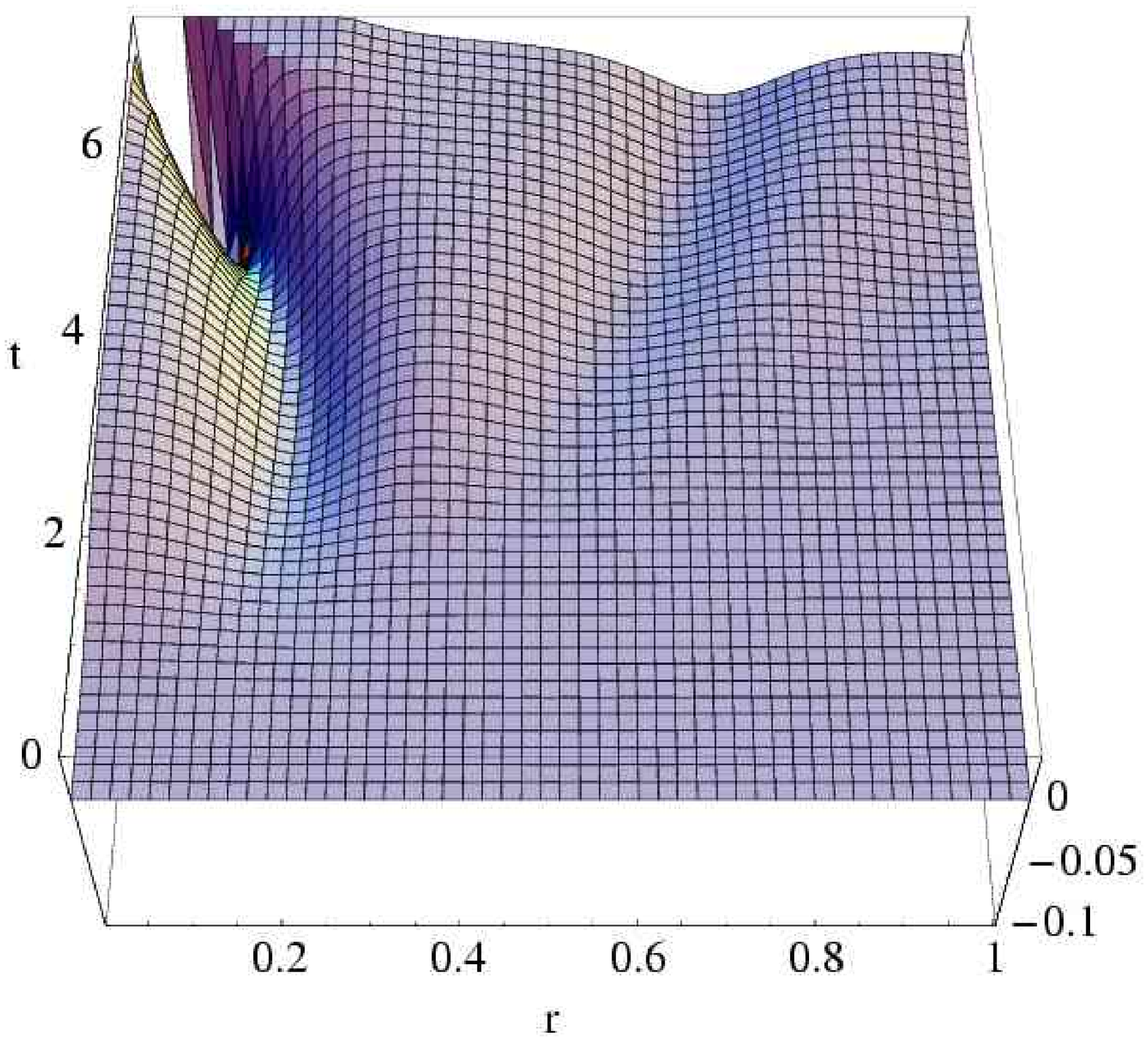,width=3.5in}}
\caption{ \label{fig:NewTerm}
 The evolution of $X-1$ in the perfectly planar symmetric case with and
 without the new term. The left figure shows the evolution with the new
 term with $\beta=0.04$, $\gamma=150$ and $\delta=0.001$. The right
 figure shows the evolution without the new term. The ghost length scale
 $L_{ghost}$ is set to be $0.01$ and the initial condition at $t=0$
 is set by $\pi=-0.2\times\exp(-x^2)$, $X-1=0$. It is evident that the
 would-be caustics at  $t\simeq 7.5$ is resolved by the new term.
}
\end{figure}

Just to show that this argument based on the Hamiltonian works, let us
consider a simple choice
%
\begin{equation}
 \tilde{F}(z) = \frac{\gamma}{2}z^2
  \left[ \tanh\left(\frac{z-\beta}{\delta}\right)+1\right],
\end{equation}
where $\beta$, $\gamma$ and $\delta$ are positive constants, and
$\tilde{F}(z)\equiv 4\sqrt{z}(\sqrt{z}F'(z))'$. We have specified
$\tilde{F}(z)$ instead of $F(z)$ just for practical purpose: since the
equation of motion includes the term 
$\tilde{F}(z)\nabla\pi\cdot\nabla z/2z$, $\tilde{F}(z)$ is easier to be
implemented in numerical codes than $F(z)$ itself. Results for the
perfectly planar symmetric case with and without the new term are shown
in Fig.~\ref{fig:NewTerm}. It is easy to see that the would-be caustics
at $t\simeq 7.5$ is resolved by the new term. 

In this subsection we considered simple models of new degrees of freedom
and new terms arising near the perfect caustics. As stated in the
beginning of this subsection, they must be taken into account since the
usual leading spatial derivative term $\nabla^4\pi$ vanishes for the
perfect caustics and the spatial derivative expansion in this background
should start differently from that in the $\pi=0$ background. We have
seen that turning on the gauge coupling of gauged ghost condensation
or/and the new term resolves the singular behavior. We must, however,
consider the simple models presented here just as an existence proof of
systems without the singular behavior. A detailed analysis of the
consistent derivative expansion and/or more realistic models of UV
completion are necessary as a future work.

\section{Bounds}
In this section, we consider the bounds on the ghost condensate.

\subsection{Jeans Instability}
\label{subsec:jeansbound}
For $M \gsim 10\MeV$, the Jeans instability time is shorter than the
lifetime of the universe, and we must consider the effects of this instability.
We have seen that the nonlinear effects dominate near interesting gravitational
sources, but the linear dynamics still controls the behavior of the system
for sufficiently weak ghostone amplitudes.
In the linear regime,
fluctuations with wavelength $\la \gsim L_{\rm J}$ grow on a
time scale
\beq
\tau \sim T_{\rm J} \frac{\la}{L_{\rm J}},
\eeq
where
\beq
L_{\rm J} \sim \frac{\MP}{M^2},
\qquad
T_{\rm J} \sim \frac{\MP^2}{M^3}
\eeq
are the Jeans length and time scales.
Wavelengths of order $L_{\rm J}$ become unstable first, and
longer wavelengths take longer to grow.
Since fluctuations on wavelength shorter than $L_{\rm J}$ are stable, we
expect the minimum size of a positive or negative energy region to be 
$L_{\rm J}$. On the other hand, the maximum size is determined by
requiring that the time scale $\tau$ above be shorter than the Hubble
time. Hence, a positive or negative region can grow within the age of
the universe if its size $L$ is in the range
\beq[range-of-L]
 L_{\rm J} \lsim L \lsim L_{\rm max},
 \label{eqn:Lrange}
\eeq
where
\beq
 L_{\rm max} \sim \frac{M}{\MP H_0} 
 \sim R_{\odot}\left(\frac{M}{100\GeV}\right).
\eeq

The unstable modes grow at least until nonlinear effects become
important. This happens for $\pi \gsim \pi_{\rm c}$, where
\beq
\pi_{\rm c} \sim \frac{\la^2}{\tau}.
\eeq
or equivalently $\Si \gsim \Si_{\rm c}$ with
\beq
\Si_{\rm c} \sim \frac{\pi_{\rm c}}{\tau}
\sim \frac{\la^2}{\tau^2}
\sim \frac{M^2}{\MP^2}.
\eeq
It is reasonable to assume that the nonlinear effects cut off the Jeans
instability at this critical amplitude.
This mechanism will fill the universe with regions of positive and negative
ghostone field with amplitude of order $\pm \Si_{\rm c}$ and the size in
the range \Eq{range-of-L}. 
Since $\Si$ is a conserved charge, there will be equal amounts of positive
and negative $\Si$.

The sun's Newtonian potential triggers the Jeans instability of the
ghost condensate and, thus, it is expected that there be a positive or
negative region around the sun. This is justified if the `aether' is
efficiently dragged by the sun and we now argue that this is indeed the
case. To do this, it is useful to work in the rest frame of the sun. Far
from the sun, the aether is moving with constant velocity $v \sim
10^{-3}$, but near the sun the velocity field will be distorted by the
presence of the sun. By using the fluid picture of the ghostone field,
we estimate the effect on a fluid particle with speed $v$ and impact
parameter $r$. The fluid particle will be a distance of order $r$ away
for a time $\De t \sim r/v$, so the change in the particle velocity in
the impulse approximation is 
\beq
\De v \sim \frac{R_{\rm S}}{r^2} \cdot \frac{r}{v} \sim \frac{R_{\rm S}}{vr},
\eeq
where $R_{\rm S}$ is the Schwarzschild radius of the source. Thus, the
change in the velocity of a fluid particle becomes comparable to or
greater than the initial velocity if $r < r_{\rm drag}$, where 
\beq
r_{\rm drag} \sim \frac{R_{\rm S}}{v^2},
\eeq
For our sun, $r_{\rm drag} \sim 10 R_{\odot}$, so the dragged
region extends \emph{outside} the solar radius.%
\footnote{This radius is still much less than the orbital radius of
Mercury.} 

We require that the absolute value of the mass of the lump with the
critical density $\rho_c$ and the size $L_{\rm max}$ be at worst less
than the solar mass: 
\beq[sun-from-bound]
 \rho_c L_{\rm max}^3 \lsim M_{\odot}.
\eeq
This requirement gives the bound 
\beq
 M \lsim 10^3\GeV.
\eeq
Since the high power of $M$ (the l.h.s. $\propto M^9$) is involved in 
\Eq{sun-from-bound}, a more stringent requirement on the mass of the
lump will not substantially improve the bound.

\subsection{Twinkling from Lensing}

We have argued that 
if $M \gsim 10\MeV$, then the Jeans instability fills the universe
with regions of positive and negative energy of size
$L \gsim L_{\rm J} \sim \MP / M^2$ with energy density
$\rho_{\rm c} \sim M^6 / \MP^2$.
This will happen everywhere, in particular in the voids between
galaxies.
Any light that travels to us from far away will therefore be
lensed by these positive and negative regions.
These positive and negative energy regions move, because the
local rest frame of the lensing regions is different from
that of our galaxy, so the result is that the observed
luminosity of any point source will change with time.
This is similar to the twinkling of the stars in the night sky caused by
time dependent temperature differences in the atmosphere. 
In this subsection, we work out the bounds on the ghost 
condensate from this effect.

Suppose that the universe is filled with regions of positive
and negative energy with size $L$ and density $\rho_{\rm c}$.
A light ray traveling through such a region will lens by an
angle
\beq
\De\th \sim \Phi \sim \frac{\rho_{\rm c} L^2}{\MP^2}
\sim \frac{M^6 L^2}{\MP^4}.
\eeq
If a light ray travels a distance $d \gg L$, then it will
undergo $N \sim d/L$ uncorrelated lensing events, so the total
angular deviation will be enhanced by a $N^{1/2}$ random walk factor:
\beq
\De\th_{\rm tot} \sim \left( \frac{d}{L} \right)^{1/2}
\frac{M^6 L^2}{\MP^4}.
\eeq
We see that the largest angular deviation comes from the largest $L$
and largest $d$.

The size of $L$ is limited by the time for the Jeans instability to
form as in (\ref{eqn:Lrange}). 
If the source is the cosmic microwave
background, then $d \sim H_0^{-1}$ and we obtain
\beq
\De\th_{\rm CMB} \sim \frac{M^{15/2}}{\MP^{11/2} H_0^2}
\sim \left( \frac{M}{100\GeV} \right)^{15/2},
\eeq
for the largest regions with the size $L\sim L_{max}$. 
The high power of $M$ makes the precise 
experimental limit on $\De\th_{\rm CMB}$
irrelevant, and we obtain the bound
\beq
M \lsim 100\GeV.
\eeq

For $M \sim 100\GeV$, the size of the largest critical region is
$L \sim 10^{12}$~cm, approximately the radius of the sun.
The local velocity of these regions relative to our galaxy is
of order $10^{-3}$, so the time scale for one of these regions to
cross the line of sight is of order a day, which is therefore the
time scale of the variation.

If there is a distant astrophysical source that is observed to shine with
very little time variation, it may give a competitive bound.
But given the high power of $M$ involved, it seems difficult to
improve on this bound significantly.


\subsection{Supernova time-delay}

Gravitational lensing considered in the previous subsection induces a 
time-delay for light-rays coming from far distances. With this
time-delay effect, observed supernovae should be older than they
appear. Thus, this effect would change the estimate of dark energy by
observation of Type Ia supernovae. Since the determination of the dark
energy by supernovae observation is known to be consistent with the WMAP
data, we require that the time-delay is sufficiently shorter than the
total time: 
\beq
 \frac{\De t}{t} \sim \left(\De\theta\right)^2
 \sim \frac{M^6L_{\rm max}^2}{\MP^4} \leq 1. 
\eeq
Note that the precise experimental limit on the $\De t/t$ is irrelevant
because of the higher power of $M$ involved in the l.h.s. 
From this we obtain the bound
\beq
 M \lsim 10^3 \GeV.
\eeq


\subsection{Nonlinear Energy Loss}

A gravitating object moving relative to the preferred frame will lose
energy.
As discussed above, the ghostone field outside a gravitating
object is in the nonlinear regime for essentially all gravitationally interesting
objects and interesting values of $M$.
In this subsection, we
estimate this energy loss in the nonlinear regime.
Interestingly, we obtain parametrically the same result as the linear
calculation, and so this is a small effect in the real world.

It is simplest to work in a frame where the ghost fluid is at rest.
As the gravitating object moves through the medium, its gravitational
potential excites the surrounding ghost fluid, and therefore the object
loses energy.
This picture is very similar to the energy loss of charged particles
in a medium, treated in standard texts on electrodynamics.
In fact, the particle picture makes the analogy very close.

Consider then a gravitating object characterized by its Schwarzschild
radius $R_{\rm S} = M_* / \MP^2$ moving with speed $v$ in the $+z$
direction relative to the preferred frame.
We estimate the effect on a particle with impact parameter $r$ using
the impulse approximation in order to see the response of the ghost
condensate in the particle picture. The moving object will be a distance
or order $r$ away for a time $\De t \sim r/v$, so the change in the
particle velocity in the impulse approximation is
\beq
\De v \sim \frac{R_{\rm S}}{r^2} \cdot \frac{r}{v} \sim \frac{R_{\rm S}}{vr}.
\eeq
In terms of the ghostone field, this means that we generate a disturbance
with $\grad\pi \sim \De v$.
To estimate the energy density
associated with this, we must remember that there is
no modification of gravity if we neglect the $\al(\grad^2 \pi)^2$ term.
The change in the ghostone field energy is therefore
\beq
\De \rho \sim \al M^2 (\grad^2 \pi)^2 \sim \frac{\al M^2 R_{\rm S}^2}{v^2 r^4}.
\eeq
The total energy transferred to the ghostone fluid is therefore
\beq
\dot{E} \sim v \frac{dE}{dz} \sim
v \myint dr\, r \frac{\al M^2 R_{\rm S}^2}{v^2 r^4}.
\eeq
This integral is dominated at small $r$, so we have
\beq
\dot{E} \sim \frac{\al M^2 R_{\rm S}^2}{v r_{\rm min}^2}.
\eeq

We now consider the short distance cutoff $r_{\rm min}$.
One cutoff comes from the use of the impulse approximation.
This is valid only if the particles moves a distance
smaller than $r$ in the time $\De t$:
\beq
a (\De t)^2 \sim \frac{R_{\rm S}}{r^2} \left( \frac{r}{v} \right)^2 \lsim r,
\eeq
which gives
\beq
r_{\rm min} \sim \frac{R_{\rm S}}{v^2}.
\eeq
For $r \sim r_{\rm min}$ we have $\De v \sim v$,
so the maximum velocity of particles with $r < r_{\rm min}$ is
$\De v \sim v$.
This shows that the contribution of particles with $r < r_{\rm min}$
is suppressed, justifying the use of $r_{\rm min}$ as a cutoff.

Another cutoff is provided by the finite size $R$ of the object itself.
This is smaller than $r_{\rm min}$ given above provided that
\beq[surfgravbound]
\frac{R_{\rm S}}{R} \gsim v^2.
\eeq
The \lhs is the surface gravity of the object, and the \rhs is of order
$10^{-6}$ for a typical astrophysical object.
For example, the sun has a surface gravity of order $10^{-5}$, while the
earth has a surface gravity of order $10^{-9}$.

Putting this together, we see that for an object of sufficiently strong
surface gravity (\Eq{surfgravbound}), the energy loss is
\beq[finalEdot]
\dot{E} \sim \al M^2 v^3.
\eeq
Note that the dependence on the mass of the object
(as well as the Planck mass) has dropped out.
\Eq{finalEdot} is the same parametric formula obtained from a linear calculation,
and completely negligible for macroscopic objects.
If the surface gravity is weak, the energy loss is even smaller.

\section{Ghost Dark Matter}
As shown in \Ref{Arkani-Hamed:2003uy} and reviewed above, the expansion of the universe
drives $\Si \to 0$.
If $\Si$ is nonzero early in the history of the universe, then the
ghostone energy density evolves with the scale factor $a$ as
\beq
\rho \sim \Si \sim a^{-3},
\eeq
with $p \sim \rho^2 / M^4 \ll \rho$ for $\rho \ll M^4$.
(Since the energy density of the universe at matter-radiation equality is
$\sim (1\eV)^4$, this requres $M \gsim 1\eV$.)
Thus, the ghostone energy and pressure redshifts just like cold dark matter.
Note that since $\Si$ is a conserved charge density, a nonzero value
of $\Si$ early in the universe corresponds to a nonzero charge asymmetry,
similar to the baryon asymmetry in the universe.
In such a scenario, the amount of dark matter is determined by very
different microphysics from the conventional freeze-out of particle dark
matter.
In particular, if we want to generate a nonzero density of $\Si$ charge
from an initial condition with no charge asymmetry (`ghostogenesis'),
the theory must contain interactions that violate the shift symmetry.
We will not discuss the required initial conditions for ghost dark matter
further, but instead turn to the viability of this scenario.

As already noted in \Ref{Arkani-Hamed:2003uy}, small
fluctuations of the ghostone field with $\de \rho / \rho \ll 1$
also behave just like cold dark matter.
This can be seen from the fact that the equation of motion
in both cases follows from the conservation of the stress-energy tensor for
linearized scalar fluctuations,
which is the same for the ghostone field as for cold dark matter.
Using the results of the present paper, we can see explicitly that the
equations of motion for the ghostone mode are equivalent to a perfect
fluid with equation of state $p = \rho^2 / 2 M^4$ even in the nonlinear regime
$\de \rho / \rho \sim 1$.

However, the ghostone fluid behaves very differently from cold dark matter
in the nonlinear regime.
To see this, let us recall the basic facts about the nonlinear dynamics of
cold dark matter.
An overdense region of cold dark matter starts gravitational collapse,
and decouples from the universe expansion.
The cold dark matter particles do not reach the center of the
overdense region because they carry angular momentum, and there is no
efficient mechanism for them to lose it.
The trajectories of cold dark matter particles generically cross, but nothing
special happens because the particles have negligible interactions.
The over-dense region therefore virializes, becoming effectively a
cloud of dark matter particles orbiting in the gravitational potential
produced by all the other dark matter particles.
At any given point in the overdense region, there are streams of dark matter
particles orbiting in many different directions.
Numerical simulations show that this kind of dark matter is quite successful in
reproducing many of the observed features of structure in our universe, although
there are open questions and possible discrepancies.
(For a recent review, see \eg \Ref{Ostriker:2003qj}.)

In the case of ghost dark matter, the trajectories of the `particles' in the
fluid picture cannot cross, since the theory is microscopically a scalar field.
The pressure term will provide a repulsive force in the over-dense regions,
and a simple argument suggests that it is more important than
higher derivative terms such as the $\al (\grad^2 \pi)^2$ term in the action.
In a region where ghostone energy is the dominant gravitational source, the
gravitational potential is given by
\beq[Phighostsource]
\grad^2 \Phi = m^2 \Si,
\eeq
where
\beq
m = \frac{M^2}{\sqrt 2 \MP} \sim \frac{1}{L_{\rm J}}.
\eeq
The equation of motion of the fluid particles is
\beq
\frac{D \vec{v}}{D t} = -\grad( \Phi + \Si),
\eeq
where the $\grad \Si$ term gives rise to the pressure force.  From
this we see that the pressure force will dominate over the
gravitational forces if $\Si \gsim \Phi$.  In the case where the
dominant source of gravity is $\Si$ (as in a halo of ghost dark
matter), \Eq{Phighostsource} shows that $\grad\Si \gsim \grad\Phi$ for
lumps smaller than $m^{-1} \sim L_{\rm J}$.  This means that the
pressure force will act to smooth out any inhomogeneities on scales
smaller than $L_{\rm J}$.

To see that the pressure force can indeed balance gravitational
forces, we can consider spherically symmetric static configurations,
\ie `ghost stars.'
The force-free condition is $\grad(\Phi + \Si) = 0$, which implies
$\Si = -\Phi + \hbox{\rm constant}$.
If the constant is nonzero, then $\Phi \sim r^2$ at large $r$,
which can be discarded because it has growing energy density at
infinity. We therefore have $\Si = -\Phi$, and \Eq{Phighostsource} gives
a solution
\beq[ghoststarsoln]
\Si = -\Phi = \Si_0 \frac{\sin (mr)}{m r},
\eeq
where $\Si_0 = \Si(r = 0)$. For this solution the $\pi$'s gradient
grows, but its contribution to the energy density is suppressed by
$M^2/M_{Pl}^2$ and thus negligible. Actually, the gradient of $\pi$ at
time $t$ after the formation of the 'ghost star' is estimated as 
\beq
 \grad^2\pi \sim m^2\dot{\pi}t \sim m^2\Sigma_0 t
   \sim m^2\frac{\sqrt{\rho_0}}{M_{Pl}}\frac{t}{t_{dyn}},
\eeq
where $t_{dyn} \sim \frac{1}{\sqrt{G_N\rho_0}}$ is the dynamical time 
for the 'ghost star' and $\rho_0=M^4\Sigma_0$ is the central
density. Thus, the energy due to the gradient is estimated as 
\beq
 \Delta\rho \sim \alpha M^2(\grad^2\pi)^2 \sim
   \alpha\frac{M^2}{M_{Pl}^2}\rho_0\left(\frac{t}{t_{dyn}}\right)^2. 
\eeq
Hence, $\Delta\rho/\rho_0$ remains suppressed by the factor
$M^2/M_{Pl}^2$ and is negligible for $t\sim t_{dyn}$. This means that
the 'ghost star' can be considered as a quasi-static configuration. 
This solution has alternating regions of positive and negative
$\Si$, and therefore does not describe a normal star.
However, the region $r < \pi / m$ has positive $\Si$ (if $\Si_0 > 0$)
and describes a region where gravity and pressure forces balance.
In fact, we can easily check that the solution in this region is stable.
The Lagrangian for fluctuations about this solution is
\beq
\scr{L} = M^4 \left[
\sfrac 12 \dot{\pi}^2 - \sfrac 12 \Si_0 \frac{\sin(mr)}{m r} 
(\grad \pi)^2 \right].
\eeq
The equation of motion for spherical perturbations is
\beq
\ddot\pi = \frac{\Si_0}{m r^2} \d_r \left[
r \sin( m r) \d_r \pi \right].
\eeq
Separating variables as $\pi = \tilde{\pi}(r)e^{-i\omega t}$,
multiplying the equation of motion by $r^2\tilde{\pi}$, integrating from
$r=0$ to $r=\pi/m$ and using integration-by-parts, we obtain 
\beq
\omega^2\int_0^{\pi/m}\left[\tilde{\pi}(r)\right]^2r^2dr = 
\frac{\Sigma_0}{m}
\int_0^{\pi/m}\left[\tilde{\pi}'(r)\right]^2r\sin(mr)dr. 
\eeq
This implies that $\omega^2>0$ unless $\tilde{\pi}'=0$ everywhere in the
interval of integration. Hence, the positive $\Sigma$ region near the
origin is stable. 
Although this solution is not realistic, it gives an explicit
example where pressure can cancel gravitational attraction.

We see that in regions where $\Si > 0$ is the dominant source of
gravity, the pressure force tends to smooth out inhomogeneities on
scales smaller than $L_{\rm J}$.
Let us ask whether the $\al (\grad^2 \pi)^2$ term is important in
these regions.
In the solution without the $\al (\grad^2 \pi)^2$ term, we have
\beq
\Si \sim \frac{\pi}{T} \sim \frac{\pi^2}{L^2}.
\eeq
From this, we can see that the $\al (\grad^2 \pi)^2$ term is
negligible for
\beq
L \gsim \frac{1}{M \sqrt{\Si}}.
\eeq
This is satisfied at $L \sim L_{\rm J}$ provided that
\beq
\label{eq:criticalamplitude}
\Si \gsim \frac{M^2}{\MP^2}.
\eeq
Note that this is the critical amplitude $\Si_{\rm c}$ discussed
in subsection \ref{subsec:jeansbound}.
If we assume that the ghostone energy density $\rho \sim M^4 \Si$
accounts for all dark matter in the universe, then \Eq{criticalamplitude}
is always satisfied in the overdense regions for $M \lsim 10$ MeV.
In that case, it is very plausible that the pressure
is sufficient to smooth out the would-be caustics that occur
during structure formation in our universe.

We emphasize that we are not claiming that the pressure prevents
caustic singularities from forming during structure formation.  For
example, a collapsing $\Si > 0$ region may `bounce' and leave behind a
region of $\Si \sim 0$ (a void) or even $\Si < 0$.  In these regions,
the singularities discussed in the previous section may become
important.  Also, a very over-dense region may lead to caustics in a
less dense region nearby.  However, even if the $\al (\grad^2 \pi)^2$
term becomes important in some small regions of the halo, leading to
negative energy regions, we argued in the previous section that the
negative energy regions tend to shrink and carry negligible energy and
they happen at very small scales.  Therefore, we do not expect them to
be important in the overall distribution of energy.

It would certainly be very interesting to perform numerical simulations
of the nonlinear dynamics of self-gravitating
ghostone halos to see whether ghost dark matter can explain
the structure of the universe we observe today.


\section{Conclusions}

Infrared modifications of gravity have potential connections to the puzzles
we are facing in cosmology such as dark energy and dark matter. Such theories
can be viewed as gravity in the Higgs phase. The simplest possibility
with a single extra degree of freedom which nonlinearly realizes the broken
time diffeomorphism is described by ghost condensation --- a scalar field
with a time dependent vacuum expectation value, $\phi \propto t$.
In a previous paper~\cite{Arkani-Hamed:2003uy}, we have shown that a healthy low-energy
effective theory can be constructed following the symmetry requirement.
It is stable for small fluctuations in the absence of gravity ($M_{\rm Pl}
\to \infty$). When gravity is turned on, there is an infrared instability
at long wavelengths and large time scale similar to the usual Jeans
instability for ordinary matter in the presence of gravity. This
potentially provides the strongest constraint on the UV cutoff of the 
effective theory. 

The leading nonlinear interactions in the effective theory of ghost
condensation are also fixed by the symmetry. By power counting, they
are irrelevant operators so they are unimportant for small
fluctuations about the ground state. However, they can become
important in the presence of large classical sources. In this paper,
we show that for realistic astrophysical gravity sources, the time
scale for the nonlinear terms to become large is shorter than the time
scale of linear propagations.  Therefore, the nonlinear effects are
dominant near these astrophysical bodies and this paper is devoted to
studying these effects.

We show that there is a simple picture for the nonlinear evolutions of
the `ghost field.' They are simply described by the geodesics in the
gravitational field of the source.  The geodesic picture necessarily
breaks down when geodesics cross and caustics form. At these points
the higher derivative terms become important.  From the numerical
simulations we see that the generic caustics can be resolved by the
leading higher derivative terms in the linear lagrangian.  However,
there exist perfect caustics which cannot be stopped by the leading
higher derivative terms. Their resolution can only be addressed by the
physics above the UV cutoff, beyond the regime of validity of the
low-energy effective theory. A potential worry is whether some
catastrophe may be caused by the negative energy region formed due to
the caustics.  We show that both analytically and in numerical
simulations that the negative energy region shrinks with the total
amount of energy inside the region staying approximately constant, so
it will not swallow the whole space. In addition, for small enough $M$
the total energy inside each of these region is extremely small by
astrophysics standards so it is not expected to give large anomalous
gravitational effects. As long as such regions can be resolved in a
full theory, we do not expect disastrous effects or strong bounds
coming from these caustics.

In fact, the nonlinear effects may even relax the bound on the UV
cutoff of the theory. In the linear approximation, we obtained a bound
of $M \lsim 10$ MeV from requiring the time scale of the Jeans
instability to be longer than the age of the universe. The nonlinear
effects can cut off the growth of the Jeans instability and hence
removes the above bound.  The total energy in a lump can only grow to
$\sim M_{\rm Pl}$ which is minuscule before the nonlinear dynamics cut
off the growth. Now the question is how the positive and negative energy
regions change the standard picture of our universe. We found that light
rays coming from far distances lens by those lumps. By requiring that 
the random walk of light rays due to the sequence of lensing events does
not smear out the observed CMB anisotropies, we found a slightly weaker
bound of $M \lsim 100 \GeV$.

Nonlinear dynamics will also play an important role if we try to
imagine that dark matter is made of ghost condensate. We have shown
that a coherent deviation of ghost condensation from the ground state
redshifts like matter, and in the linear regime the growth of the
perturbations is identical to the usual cold dark matter. Whether it
can give rise to the correct structure formation in the universe
depends on the nonlinear evolution. This is an interesting but
complicated question, which may require detailed simulations.  It is
left for future studies.

In this paper, we have begun the investigation of the nonlinear
dynamics of ghost condensation as a Higgs phase of gravity, which
exhibits a rich structure. While we have obtained a qualitative
picture of some of the dynamics, a more complete understanding is
clearly desirable and deserves further investigations, as required for
making more concrete connections with various cosmological questions.

\section*{Acknowledgments}
We would like to thank Paolo Creminelli, Lev Kofman and Matias
Zaldarriaga for useful discussions.  S.~M. thanks Aspen Center for
Physics where part of the work was done.  N.~A-H. is supported a David
and Lucille Packard Foundation Fellowship for Science and Engineering,
and by the DOE under contract DE-FG02-91ER40654.  H.-C.~C. and
T.~W. are supported by NSF grant PHY-0244821, M.~A.~L. is supported by
NSF grant PHY-009954.

\appendix{Appendix: Higgs Phase of Full General Relativity}
\subsection{Fluid Picture}
We now show that the fluid picture derived in the main text holds in 
full general relativity, \ie allowing relativistic motion and strong
gravitational fields.
In order to remain within the realm of validity of the effective theory,
we must restrict ourselves to small ghostone ampltiudes, \ie
\beq
\Si \ll 1.
\eeq
Neglecting the $\al$ term in the action, the equations of motion of
the ghostone field can be written
\beq[eomfullGR]
\nabla_\mu J^\mu = 0,
\eeq
where
\beq[JfullGR]
J^\mu = \Si u^\mu,
\eeq
and
\beq
u^\mu = \nabla^\mu \phi.
\eeq
We cannot interpret $u^\mu$ as a 4-velocity because $u^2 = X \ne 1$ in
general.
However, this effect is $\scr{O}(\Si^2)$, as can be seen by defining a
normalized 4-vector $\hat{u}_\mu$ by
\beq
\hat{u}^\mu = \frac{u^\mu}{\sqrt{X}} = u^\mu \left[ 1 + \scr{O}(\Si^2) \right].
\eeq
We can therefore interpret
\beq
u^\mu = (\dot\phi, -\grad\phi)
\eeq
as the 4-velocity of the fluid,
up to $\scr{O}(\Si^2)$ corrections.

It now remains only to satisfy the relation between $\Si$ and $\phi$.
Differentiating the relation
\beq
g^{\mu\nu} \nabla_\mu \phi \nabla_\nu \phi = X = 1 + 2\Si,
\eeq
we obtain
\beq
\nabla_\mu\Si = \sfrac 12 \nabla_\mu ( \nabla^\rho \phi \nabla_\rho \phi)
= \nabla^\rho \phi \nabla_\mu \nabla_\rho \phi
= \nabla^\rho \phi \nabla_\rho \nabla_\mu \phi
\eeq
or
\beq[worldlineeom]
u^\rho \nabla_\rho u_\mu = \nabla_\mu \Si.
\eeq
This has the form of the geodesic equation of a particle subject to an
additional gradient force with potential $\Si$.
For the special case $\Si = 0$, the fluid particles follow geodesics.
The velocity field satisfies the curl-free condition
\beq[curlfree]
\nabla_{[\mu} u_{\nu]} = 0,
\eeq
implying that the flow is irrotational.

Conversely, any family of particle trajectories satisfying \Eq{worldlineeom}
with initial conditions 
\beq
u^\mu | = \hbox{\rm given},
\qquad
\nabla_{[\mu} u_{\nu ]} | = 0
\eeq
on some initial value hypersurface
defines a solution of the ghostone equations of motion.
To see this, note that the
vanishing of $\nabla_{[\mu} u_{\nu]}$ is preserved by
the equation of motion \Eq{worldlineeom}.
This can be seen from the identity
\beq[shinid]
u^\rho \nab\rho (\nabla_{[\mu} u_{\nu]})
= (\nabla_{[\nu} u_{\rho]}) \nab\mu u^\rho
- (\nabla_{[\mu} u_{\rho]}) \nab\nu u^\rho,
\eeq
which is derived as follows:
\beq
u^\rho \nabla_\rho \nabla_\mu u_\nu &=
u^\rho (\nabla_\mu \nabla_\rho u_\nu - R_{\rho\mu\nu}{}^\si)
\nonumber
\\
&= -\nabla_\mu u^\rho \nabla_\rho u_\nu + \nabla_\mu(u^\rho \nabla_\rho u_\nu)
- R_{\rho\mu\nu\si} u^\rho u^\si
\nonumber
\\
&= -\nabla_\mu u^\rho \nabla_{[\rho} u_{\nu]}
- \nabla_\mu u^\rho \nabla_\nu u_\rho
+ \nabla_\mu \nabla_\nu \Si 
- R_{\rho\mu\nu\si} u^\rho u^\si.
\eeq
Antisymmetrizing in $\mu$ and $\nu$ then gives \Eq{shinid}.
Therefore, the velocity field $u_\mu$ can be written as a
gradient $u_\mu = \nabla_\mu\phi$.
The field $\phi$ now obeys the equation of motion \Eq{eomfullGR},
with
\beq
g^{\mu\nu} \nabla_\mu \phi \nabla_\nu \phi = X + \hbox{\rm constant}.
\eeq
The integration constant can be seen to vanish by imposing appropriate
boundary conditions at infinity.

This means that many of the results derived in the main text hold in full
general relativity.
In particular, in regions where the solution is given by $\Si = 0$,
the local preferred frame given by $\phi$ is a freely falling frame.
In this sense, the aether is `dragged' even by strong gravitational fields.

\subsection{Negative Energy Region Shrinks}
\label{sec:shrink}
We now show that any $\Si < 0$ region shrinks in full general relativity,
neglecting the $\al (\grad^2 \pi)^2$ term in the equation of motion.

Now let us consider a hypersurface $\Sigma=0$ defining the boundary
between a negative energy region ($\Sigma<0$) and the surrounding
positive energy region ($\Sigma>0$).
On the boundary $\Sigma=0$, the
equation of motion is reduced to
%
\begin{equation}
 \partial_{\mu}\Sigma u^{\mu} = 0.
\end{equation}
Now let us calculate the acceleration
$a^{\mu}=u^{\rho}\nabla_{\rho}u^{\mu}$.
By differentiating the definition of $\Sigma$, we obtain
%
\begin{equation}
 \partial_{\mu}\Sigma = u^{\rho}\nabla_{\mu}u_{\rho}
  = u^{\rho}\nabla_{\rho}u_{\mu},
\end{equation}
where we have used $\nabla_{[\mu}u_{\rho]}=0$, and thus
%
\begin{equation}
 a^{\mu} = \partial^{\mu}\Sigma.
\end{equation}
With this expression it is easy to know the direction to which the
boundary hypersurface evolves:
%
\begin{equation}
 a^{\mu}\nabla_{\mu}\Sigma = \partial^{\mu}\Sigma\partial_{\mu}\Sigma
  < 0.
\end{equation}
Namely, the boundary hypersurface evolves towards the region with
negative $\Sigma$. In other words, the negative energy region
shrinks.

\subsection{Relativistic Caustic}
We now show how the caustic solutions discussed in
subsection \ref{subsec:causticsoln} can be found in an elegant way
in the relativistic formulation.
We consider the case of $1+1$ dimensions, which is relevant for the
generic caustic.
We have seen that the spacetime surface $\phi(x) = 0$ 
is perpendicular to the family of geodesics.
We can therefore use Gaussian normal coordinates based on this surface.
In this coordinate system, $\phi$ is the `time' variable, and the
metric in the $(x, t)$ subspace is
\beq[GNmetric]
ds^2 = d\phi^2 - A^2(\phi, \si) d\si^2,
\eeq
where the spatial variable $\si$ parameterizes the position along the
$\phi = {\rm constant}$ surfaces.
Without loss of generality, we can choose $A(0,\si) = {\rm constant}$,
since this corresponds to choosing the `space' coordinate on the surface
$\phi = 0$ to be proper distance.
The geodesics perpendicular to the surfaces of constant $\phi$
are simply $\si = {\rm constant}$ in these coordinates.


The condition that the metric \Eq{GNmetric} is flat is
\beq
\d_\phi^2 A = 0.
\eeq
(For example, one can check that this guarantees the vanishing of the
Riemann tensor.)
Imposing $A(0,\si) = 1$, the general solution is
\beq
A(\phi, \si) = \left[ 1 - \phi / \pc(\si) \right],
\eeq
where 
$\pc(\si)$ is a function that specifies the
distribution of velocities on the surface $\phi = 0$.
Note that the coordinate system breaks down when $\phi = \pc(\si)$;
this is a sign of the caustic singularity.
For a caustic singularity 

In this coordinate system, it is trivial to compute invariants such as
\beq
\Box\phi &= \frac{1}{\sqrt{-g}} \d_\mu \left[ \sqrt{-g}\, g^{\mu\nu}
\d_\nu \phi \right]
= \frac{1}{A} \d_\phi A
\\
&= \frac{1}{\tau},
\eeq
where
\beq
\tau = \phi - \pc
\eeq
is the time to the caustic.
Similarly,
\beq
\Box^2\phi &= \frac{1}{\sqrt{-g}} \d_\mu \left[ \sqrt{-g}\, g^{\mu\nu}
\d_\nu \Box \phi \right]
= \frac{1}{A} \d_\phi \left[ A \d_\phi \Box\phi \right]
- \frac{1}{A} \d_\si \left[ A^{-1} \d_\si \Box \phi \right]
\\
&= -\frac{1}{\tau^3} -\frac{\pc \left[ (\pc')^2
+ \pc\pc^{\prime\prime} \right]}{\tau^4}
+ \frac{3 \pc^2 (\pc')^2}{\tau^5}.
\eeq
Expanding the function $\pc(\si)$ for small $\si$, we obtain
\beq
\Box^2\phi \sim -\frac{1}{\tau^3}
- \frac{\pc^2(0) \pc''(0)}{\tau^4}
+ \scr{O}(\si^2).
\eeq
This agrees precisely with the results of subsection
\ref{subsec:causticsoln}
in the nonrelativistic limit,
with the identifications
$\si = x_0$, $\pc(\si) = t_{\rm c}$, and $\tau = t - T$.


\end{document}